\documentclass[12pt]{article}

\usepackage[nosort]{cite}
\usepackage{epsfig}
\usepackage{amsmath}
\usepackage{amsfonts}
\usepackage{amssymb}
\usepackage{multirow}
\usepackage{array}
\usepackage{subfigure}

\usepackage[affil-it]{authblk}

\title{\bf Z-peaked excess 
from heavy gluon decays to vector-like quarks}
\author{Natascia Vignaroli}
\affil{\small Department of Physics and Astronomy, \\
Michigan State University, East Lansing 48824, USA}

\begin{document}
\maketitle

 
 
\begin{abstract}
A 3 sigma excess has been recently announced by ATLAS in events with Z-peaked dilepton pairs, jets, and large transverse missing energy. We interpret this finding 
in the context of composite Higgs / RS theories. 
We find
that composite Higgs theories with custodial symmetry protection
to the $Zb\bar{b}$ coupling predict a significant contribution to $ZZbb$ (and to $hhbb$) final states coming from heavy gluon decays to pairs of bottom-partner vectorlike quarks. The heavy gluon to vector-like quarks signal is largely accepted by the ATLAS selection if one of the $Z$ boson in the $ZZbb$ final state decays leptonically and the other to neutrinos. 
For a bottom partner of $\sim$900 GeV, we find that the ATLAS excess can be reproduced by composite Higgs models,
in an experimentally allowed parameter space, for heavy gluon masses roughly in a range 1.87 - 2.15 TeV and for heavy gluon couplings to light quarks within $\sim(0.3-0.65) g_S$. We briefly discuss the implication of this result for future experimental tests.
\end{abstract}

\newpage

\section{Introduction}
An excess above the expected number of Standard Model events, with a statistical significance of three standard deviations, has been recently measured by ATLAS 
in events containing a same-flavor opposite-sign dilepton pair, jets, and large transverse missing energy \cite{Aad:2015wqa}. The excess is only found in the on-Z region, for dilepton mass near the Z peak. \footnote{A similar analysis in the same channel considered by ATLAS has been also performed by CMS \cite{Khachatryan:2015lwa}. Probably due to the different kinematic selection, as also speculated in \cite{Aad:2015wqa} (in particular CMS does not apply a cut in $H_T$), CMS does not find any excess of events above the background in the on-Z region. On the other hand CMS claims a 2.6 $\sigma$ excess in the off-Z region, where no deviation from SM expectations are observed by ATLAS. In our study we will not try to explain the CMS excess and we will not analyze the CMS results.}\\
The on-Z ATLAS search is focused on generalized gauge mediated (GGM) supersymmetry breaking models and aims at selecting a signature where pair-produced gluinos decay via neutralinos to a gravitino plus a Z boson. 
Two examples of GGM models are considered by ATLAS to interpret the results.
Recent studies \cite{Barenboim:2015afa,1504.02244,1504.02752} have also analyzed the ATLAS results in the context of supersymmetry. In \cite{Barenboim:2015afa} it is found that ``only relatively light gluinos, together with a heavy neutralino decaying predominantly to Z boson plus a light gravitino could reproduce the excess", while \cite{1504.02244} shows that GGM models are unlikely to reproduce the ATLAS results and that alternative scenarios with two massive neutralinos, which can be possibly realized in NMSSM, are favored. The study in \cite{1504.02752} arrives at a similar conclusion and states that, after combining the relevant constraints from LHC searches for new Physics, GGM models cannot explain the ATLAS excess. \\
In this paper we present an interpretation of the ATLAS results in the context of composite Higgs \cite{Kaplan:1983fs}/ RS \cite{Randall:1999ee} theories. 
Composite Higgs models are compelling theories to solve the hierarchy problem and to explain the electroweak symmetry breaking. In the most natural parameter region of the theory \cite{Marzocca:2012zn, Panico:2012uw}, new vector resonances are above the threshold for decays into new vector-like top partners and we will show that the ATLAS excess can indeed be produced by a heavy gluon resonance decaying to vector-like quarks. In particular, we will show that in model with a custodial symmetry protecting the $Zbb$ coupling \cite{Agashe:2006at} heavy gluon decays to bottom partners can give a large contribution to $hhbb$ or $ZZbb$ final states. In the latter case, If one of the Z decays leptonically and the other to neutrinos, thus producing large missing energy, the heavy gluon to vector-like quarks signal is largely accepted by the ATLAS selection \cite{Aad:2015wqa}.  \\
We will analyze in detail the heavy gluon to bottom partners to $ZZbb$ signature, considering as a benchmark a concrete and experimentally allowed realization of the composite higgs theory.
We will apply the ATLAS selection to Monte Carlo generated samples of the heavy gluon signal, finding the region of the model parameter space, currently not excluded by experiments, which can reproduce the ATLAS excess.\\

The paper is organized as follows: we present our benchmark model in sec. \ref{sec:model}; we discuss the experimental limits on the model in sec. \ref{sec:excl}; we apply the ATLAS selection to the composite Higgs signal in sec. \ref{sec:atlas} and present the results in sec. \ref{sec:results}. We finally draw our conclusions and discuss the implication of our findings, especially for future experimental tests, in sec. \ref{sec:discussion}.


\section{The model }\label{sec:model} 

We evaluate the heavy gluon contribution to the signal examined by ATLAS in a concrete model.    
We consider the effective description of heavy gluon and vector-like quarks (vlq) interactions derived in \cite{Bini:2011zb, Vignaroli:2012si}. This effective model can reproduce the low energy limit of compelling theories to solve the hierarchy problem and explain the electroweak symmetry breaking (EWSB), such as composite Higgs \cite{Agashe:2004rs} or RS in a dual five dimensional picture \cite{Agashe:2003zs}. \\
The effective theory consists of two sectors, the elementary sector with elementary particles and (analogous to) Standard Model (SM) gauge symmetries and the composite sector, with composite particles, resulting from a new strong dynamics, which include the Higgs and new fermionic and vectorial resonances. The two sectors mix with each other. In particular, a new composite gluon associated with a $SU(3)^{comp}_c$ gauge symmetry in the strong sector mixes with the elementary gluon of $SU(3)^{ele}_c$. After diagonalizing the mixing, the physical states consist of the massless SM gluon and of a new heavy gluon $G^{*}$. The $G^{*}$ interactions with SM particles and new vlqs from the composite sector are controlled by the parameter which describes the rotation of the elementary-composite mixing to the physical basis:
\begin{equation}\label{eq:tg3}
\tan\theta_{3}=\frac{g^{ele}_{3}}{g^{comp}_{3}}<1 \qquad g_S = g^{ele}_{3} \cos\theta_{3}=g^{comp}_{3} \sin\theta_{3} \ ,
\end{equation}
  where $g_S$ is the QCD coupling and $g^{ele}_{3}$ ($g^{comp}_{3}$) is the $SU(3)^{ele}_c$ ($SU(3)^{comp}_c$) coupling. It is assumed that $g^{comp}_{3}> g^{ele}_{3}$.\\

New vlqs also emerge from the composite sector. 
Based on the minimal $SO(5)/SO(4)$ coset for composite Higgs models \cite{Agashe:2004rs}, which includes a custodial symmetry protection against large correction to the $\rho$ parameter and to the $Zb_L\bar{b}_L$ coupling \cite{Agashe:2006at}, we consider composite fermions in fundamental representations of $SO(5)$. $\mathbf{ 5}_{X}$ representations decompose as $ ( \mathbf{ 2,2})_{X} \oplus (\mathbf{1,1})_X$ under $SO(4)\times U(1)_X \sim SU(2)_L\times SU(2)_R\times U(1)_X$ -- $U(1)_X$ is introduced to correctly reproduce the hypercharge, $X=Y-T_{3R}$. The composite fermion content we consider is thus
\begin{align}
\begin{split}
& \mathcal{Q}=\left( \begin{array}{cc} T & T_{5/3} \\ B & T_{2/3}\end{array} \right)=(\mathbf{3, 2,2})_{2/3} \qquad \tilde{T}=(\mathbf{3, 1,1})_{2/3} \\
& \mathcal{Q'}=\left( \begin{array}{cc} B_{-1/3} & T'  \\ B_{-4/3} & B'\end{array} \right)=(\mathbf{3, 2,2})_{-1/3} \qquad \tilde{B}=(\mathbf{3, 1,1})_{-1/3} \ ,
\end{split}
\end{align}
where we have specified the quantum numbers under $SU(3)_c \times SU(2)_L\times SU(2)_R\times U(1)_X$.
Similarly to the mixing for the vector fields, composite fermions also mix, via linear mass mixing terms \cite{Kaplan:1991dc}, with the top and bottom quarks of the elementary sector. In particular, the doublet $q^3_L=(t^{ele}_L , b^{ele}_L)$ mixes with the $(T,B)$ $SU(2)_L$ doublet in the $\mathbf{5}_{2/3}$ 
and has a weak mixing with the $(T',B')$ doublet in the $\mathbf{5}_{-1/3}$; $t^{ele}_R$, $b^{ele}_R$ mix respectively with the electroweak singlets $\tilde{T}$ and $\tilde{B}$. The Lagrangian of the mixing in the elementary-composite basis reads:
\begin{equation}\label{eq:mix}
\mathcal{L}^{mix} = -\Delta_{L1} \bar{q}^3_L (T, B) - \Delta_{L2} \bar{q}^3_L (T', B') -\Delta_{R1} \bar{t}_R \tilde{T} -\Delta_{R2} \bar{b}_R \tilde{B} + \text{H.c.}\ ,
\end{equation}
where $\Delta_{L2}\ll \Delta_{L1}\sim \Delta_{R1}\sim \Delta_{R2}$. The mixing leads to a scenario of partial compositeness of the top and bottom SM particles, which become superpositions of their composite and elementary modes and acquire their masses through the interactions of their composite modes with the composite Higgs:  
\begin{equation}\label{eq: mass}
m_t \simeq Y_{*} s_L s_R \frac{v}{\sqrt{2}} \qquad m_b \simeq Y_{*} s_2 s_{bR} \frac{v}{\sqrt{2}} \ ,
\end{equation}
where $Y_{*}$ is the Yukawa coupling among composites, of $\mathcal{O}(1)$, and $s_L$, $s_R$, $s_{bR}$ represent respectively the $q^3_L$, the $t_R$ and the $b_{R}$ degree of compositeness 
\footnote{$s_L$, $s_R$, $s_{bR}$ are related to the mixing parameters in (\ref{eq:mix}) by:
\[
s_L = \frac{\Delta_{L1}}{\sqrt{\Delta^2_{L1}+M^2_{\mathcal{Q}}}} \qquad s_R = \frac{\Delta_{R1}}{\sqrt{\Delta^2_{R1}+M^2_{\tilde{T}}}} \qquad s_{bR} = \frac{\Delta_{R2}}{\sqrt{\Delta^2_{R2}+M^2_{\tilde{B}}}} \ ,
\]
where $M_{\mathcal{Q}}$, $M_{\tilde{T}}$, $M_{\tilde{B}}$ are the $\mathcal{Q}$, $\tilde{T}$ and $\tilde{B}$ masses before the elementary-composite mixing.
}. $s_2 \ll 1$ is a parameter associated with the $q^3_L$ mixing with the $(T', B')$ $SU(2)$ doublet in the $\mathbf{5}_{-1/3}$ second tower of resonance, which permits the generation of the bottom mass. The weakness of this mixing justifies the smallness of the $m_b/m_t$ ratio and preserves the custodial symmetry protection to the $Zb_L \bar{b}_L$ coupling. We refer the reader to \cite{Bini:2011zb, Vignaroli:2012si} for more details on the model.\\

In our study, for a reason which will appear clear in the next paragraphs, we are particularly interested in the bottom partners of the $\mathbf{5}_{-1/3}$ and we will consider the following simplifying assumptions on the vlq spectrum:
\begin{itemize}
\item $M_{\tilde{B}}\gg M_{\mathcal{Q'}}$. The $\tilde{B}$ decay modes depend on the details of the electroweak mixings among bottom fermions \cite{Bini:2011zb,Chala:2013ega}. We thus decide to consider a decoupled electroweak singlet bottom-prime to reduce the model dependence of our analysis.
\item $M_{\mathcal{Q}}\gg M_{\mathcal{Q'}}$. This is a simplifying conservative choice, since some of the heavy quarks in the $\mathcal{Q}$, in particular the $T$ and the $T_{2/3}$, which decay $\simeq$50\% into $Zt$, can also contribute to the excess of events measured by ATLAS.
\end{itemize}

As an effect of the custodial symmetry, which protects the $Zb\bar{b}$ coupling and demands an $SU(2)_L\times SU(2)_R$ bidoublet representation for the $\mathcal{Q'}$ composite fermions, a degenerate doublet of bottom partners, the $B_{-1/3}$ and the $B'$, with the same mass and Yukawa coupling (before the EWSB), appears in the spectrum. We can rotate the degenerate states to the new fields \cite{Chala:2013ega}: 
\begin{equation}\label{eq:Bz-BH}
B_H = \frac{1}{\sqrt{2}}(B_{-1/3}+B') \qquad B_Z = \frac{1}{\sqrt{2}}(B_{-1/3}-B') \ ,
\end{equation} 
which will coincide with the mass eigenstates (up to a negligible correction for the $B_H$, coming from electroweak mixing effects). The $B_H$ and the $B_Z$ vlqs completely decay respectively into $bh$ and into $bZ$ (more details can be found in \cite{Chala:2013ega, Atre:2008iu} and in the appendix): 
\begin{equation}\label{eq:Bz-BH-br}
BR(B_H\to hb) = BR(B_Z\to Zb)=1 .
\end{equation} 
Notice that If we had a single $SU(2)_L$ doublet $(T', B')$, we would have had a single bottom-prime quark, the $B'$, decaying 50\% to $Zb$ and 50\% to $hb$, as explained in the appendix. In this case, the $G^{*}$ decays to $B'$ pairs would have produced a mixed $Zbhb$ final state 50\% of the time. The custodial symmetry, which leads to the $B_H$ ($B_Z$) mass states fully decaying into $hb$ ($Zb$) prevents the production of a $Zbhb$ final state via the $G^{*}$ decays into bottom-prime pairs\footnote{except possible contributions from the $G^{*}$ decays into $\tilde{B}$ pairs. 
}. This implies that the cross sections for the $ZZbb$ and the $hhbb$ final states are enhanced by a factor of two compared to the non-custodial case\footnote{assuming the same branching ratio for the heavy gluon decays to a generic pair of bottom-primes in the custodial and non-custodial cases.}.\\
The rest of the non-decoupled vlqs have the following decay patterns:
\begin{align} \label{eq:BR}
\begin{split}
& BR(B_{-4/3} \to W^{-} b)= BR(T' \to W^{+} b) =1 \\
& BR(\tilde{T} \to Wb) =0.5 \qquad BR(\tilde{T} \to Zt) = BR(\tilde{T} \to ht) =0.25
\end{split}
\end{align}
\noindent
In addition to the elementary-composite mixing, the EWSB induces a further mixing among same charge quarks. After the electroweak mixing diagonalization (the mass matrices can be found in \cite{Vignaroli:2011um}), we finally arrive at the mass basis. We select the following set of mixing parameters, for which the bottom and top masses are correctly reproduced, and which give vlq (physical) masses that fulfill the limits from direct searches for top-partners at the LHC: 
\begin{align}\label{eq:param}
\begin{split}
& Y_{*} =3 \qquad s_L =0.57 \qquad s_R = 0.6 \qquad s_{bR}=0.3 
\end{split}
\end{align}
\footnote{The remaining parameters, which we do not report in eq. (\ref{eq:param}) because are less relevant for the $G^{*}$ phenomenology, are: $s_2=s_3=0.03$, $s_4=0.05$ (where $s_3$, $s_4$ are small mixing parameters proportional to $s_2$ \cite{Vignaroli:2012si}).} 
The bare masses for $\tilde{T}$ and for the $\mathcal{Q'}$ bidoublet have been fixed respectively at 880 GeV and 930 GeV. The physical masses are: 
\begin{center}
$M_{B_{-4/3}}$ = 930 GeV \  [912 GeV \cite{CMS:2014dka}] \hspace{0.5cm} $M_{T'}$ = 945 GeV \ [912 GeV \cite{CMS:2014dka}] \\ \vspace{0.5cm} 
$M_{\tilde{T}}$ = 900 GeV \ [800 GeV \cite{CMS:2014dka, atlas-conf-2015-012}] \hspace{0.5cm}
$M_{B_{H}}$ = 955 GeV \  [846 GeV \cite{CMS:2014bfa}] \\ \vspace{0.5cm} $M_{B_Z}$ = 930 GeV \ [700 GeV \cite{CMS:2013zea}] 
\end{center}
We have indicated in parenthesis, for each vectror-like quark, the strongest mass limit placed by LHC and a reference to the corresponding search.  \\

\begin{figure}
\centering
\includegraphics[width=0.6\textwidth]{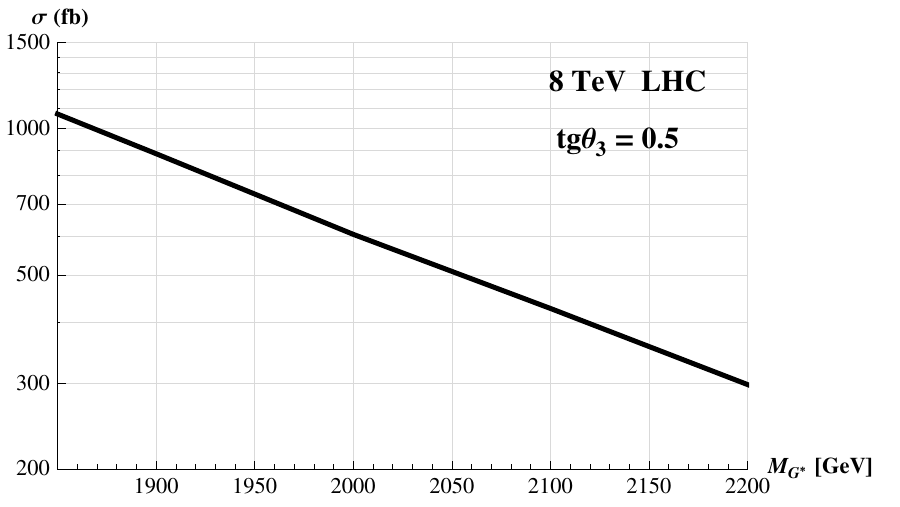}
\caption{$G^{*}$ production cross section at the $\sqrt{s}=8$ TeV LHC, for a value $\tan\theta_3=0.5$. The cross section scales with $\tan\theta_3$ as $\tan^2\theta_3$.}
\label{fig:Gstar-xsec}
\end{figure}

After having discussed the fermionic spectrum we now briefly examine the $G^{*}$ phenomenology. As anticipated, the heavy gluon interactions are ruled by the $\tan\theta_3$ parameter in (\ref{eq:tg3}). The heavy gluon is essentially a composite particle, which thus interacts strongly with the composite modes. In particular, the coupling of the interactions with elementary modes, thus with light quarks, is $g_S \tan\theta_{3}$, that with composite modes, as the $B_Z$ and $B_H$ bottom partners, is $g_S \cot\theta_{3}$. The $G^{*}$ branching ratios read \cite{Bini:2011zb}:
\begin{align}
\begin{split}\label{eq:BR-Gstar}
& BR(G^{*} \to jj) = \frac{2 }{3} \alpha_S \tan^2\theta_{3} M_{G^{*}} \\
& BR(G^{*} \to t_L \bar{t}_L + b_L \bar{b}_L) = \frac{\alpha_S }{6} \left( s^2_L \cot\theta_{3} - c^2_L \tan\theta_{3} \right)^2 M_{G^{*}} \\
& BR(G^{*} \to t_R \bar{t}_R) =\frac{\alpha_S}{12} \left( s^2_R \cot\theta_{3} - c^2_R \tan\theta_{3} \right)^2 M_{G^{*}} \\
& BR(G^{*} \to b_R \bar{b}_R) = \frac{\alpha_S}{12} \left( s^2_{bR} \cot\theta_{3} - c^2_{bR} \tan\theta_{3} \right)^2 M_{G^{*}} \\
& BR(G^{*} \to \tilde{T}\bar{t} +\bar{\tilde{T}}t ) =\frac{\alpha_S}{6} M_{G^{*}} \frac{s^2_R c^2_R}{\sin^2\theta_3 \cos^2\theta_3} \left(1- \frac{M^2_{\tilde{T}}}{M^2_{G^{*}}} \right)\left( 1- \frac{1}{2}\frac{M^2_{\tilde{T}}}{M^2_{G^{*}}} - \frac{1}{2}\frac{M^4_{\tilde{T}}}{M^4_{G^{*}}} \right) \\
&  BR(G^{*} \to \bar{\tilde{T}} \tilde{T})=\frac{\alpha_S}{12} M_{G^{*}} \left\{ \left[ \left( c^2_R \cot\theta_3 - s^2_R \tan\theta_3 \right)^2 +\cot^2\theta_3 \right] \left( 1- \frac{M^2_{\tilde{T}}}{M^2_{G^{*}}}\right) \right. \\
& \hspace{3cm} \left.+6 \left( c^2_R \cot^2\theta_3 - s^2_R \right)\frac{M^2_{\tilde{T}}}{M^2_{G^{*}}} \right\} \sqrt{1-4\frac{M^2_{\tilde{T}}}{M^2_{G^{*}}}}\\
& BR(G^{*} \to \bar{Q'} Q') =\frac{\alpha_S}{6} M_{G^{*}} \cot^2\theta_3 \left( 1+2 \frac{M^2_{Q'}}{M^2_{G^{*}}}\right)\sqrt{1-4\frac{M^2_{Q'}}{M^2_{G^{*}}}}
\end{split}
\end{align}
where, $c_x=\sqrt{1-s^2_x}$, $x=L,R,bR$ and $Q'$ denotes a generic resonance in the $\mathcal{Q'}$ bi-doublet: $Q'=B_{Z}, B_{H}, B_{-1/4}, T'$. \\
As an effect of the strong interaction with the heavy vlqs, $G^{*}$ decays predominantly into pairs of vlqs above the threshold 2$M_{vlq}$. In particular, in the $G^{*}$ mass region relevant to our analysis, we find
\begin{equation}
 BR(G^{*} \to B_Z \bar{B}_Z) \simeq 0.18 \ .
 \end{equation}
The $G^{*}$ is produced by quark-antiquark annihilation (the gluon fusion production is suppressed by gauge invariance). The production cross section, due to the $g_S \tan\theta_3$ coupling of the $G^{*}$ interactions with light quarks, depends quadratically on $\tan\theta_3$. We show in fig. \ref{fig:Gstar-xsec} the $G^{*}$ production cross section at the $\sqrt{s}=8$ TeV LHC, for a value $\tan\theta_3=0.5$.\\

\begin{figure}
\centering
\includegraphics[width=0.6\textwidth]{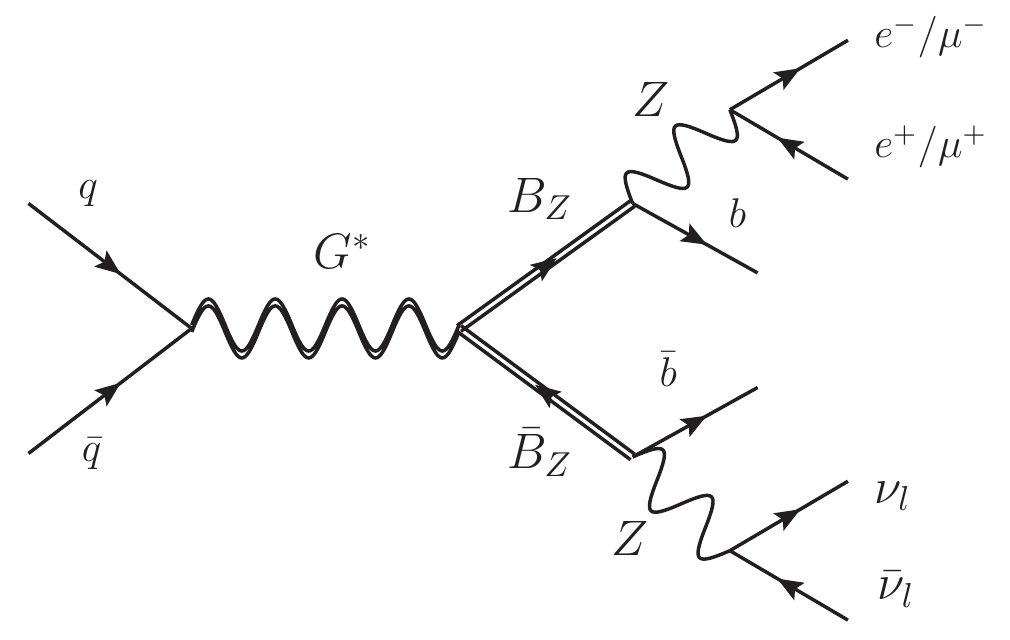}
\caption{\small Feynman diagram for the heavy gluon decays to $B_Z$ vector-like quark pairs, leading to a signature with two same flavor opposite charge leptons plus jets and missing energy.}
\label{fig:signal}
\end{figure}

Summarizing the main results of this section, we have found that minimal composite Higgs theories with custodial symmetry protection to the $\rho$ parameter and to the $Zb\bar{b}$ coupling and which can generate both the top and the bottom masses, predict a significant contribution to $hhbb$ and $ZZbb$ final states (enhanced by a factor of 2 compared to the non-custodial case) coming from heavy gluon decays to bottom partners. In particular, the process $pp \to G^{*} \to B_Z \bar{B}_Z \to ZZb\bar{b}$ will be the focus of our analysis. If one of the $Z$ of the final state decays to letpons and the other to neutrinos, as shown in fig. \ref{fig:signal}, we can indeed have a significant contribution to the excess of events measured by ATLAS\footnote{ The contribution from $G^{*} \to \tilde{T} \bar{\tilde{T}} \to ZZ t \bar{t}$ to the ATLAS signal is found to be small, below the 5\% of the total contribution, and will be thus neglected in our analysis. }.

\section{Current limits from direct searches for heavy gluon resonances}\label{sec:excl}

In this section we discuss the portion of the $G^{*}$ mass {\it vs} coupling parameter space which has been excluded by current searches at the LHC. There have been no dedicated searches so far for heavy gluon decays to vector-like quarks and the only test at colliders for $G^{*}$ theories 
can be obtained from the searches for dijet and $t\bar{t}$ resonances. We will see that, since the $G^{*}$ branching fraction for decays to dijet or $t\bar{t}$ are small when the heavy gluon is above the threshold for decaying into pairs of vector-like quarks, the searches in the dijet and $t\bar{t}$ channels leave a large part of the model parameter space untested. \\
The strongest constraints on $t\bar{t}$ resonances come from a recent CMS combined analysis of the searches in different final states \cite{CMS:2015nza}. We use the upper limits in \cite{CMS:2015nza} on the production cross section times $t \bar{t}$ branching ratio of a Kaluza-Klein gluon to extract the exclusion region for the $G^{*}$ in the $(M_{G^{*}}, \tan\theta_{3})$ plane. We find that searches in the $t\bar{t}$ channel can only exclude a portion of the parameter space at lower $G^{*}$ masses, below $\sim$1.87 TeV. The small region of the parameter space excluded by $t \bar{t}$ searches which is of interest to our study is colored brown in our final plot of fig. \ref{fig:results}.  \\
Searches for new particles in the dijet mass spectrum generally apply to narrow resonances with an intrinsic width much smaller than the experimental dijet mass resolution and cannot place limits on wide resonances like a heavy gluon decaying to top partners. 
CMS has however recently extended his search in the dijet channel to include an analysis for wide dijet resonances \cite{Khachatryan:2015sja}. We use the results of this study to extract the limits on the $G^{*}$ parameter space. In particular we use the upper limit, quoted in \cite{Khachatryan:2015sja}, on the production cross section times dijet branching ratio times acceptance ($\sigma \times BR \times A$) of a generic $q\bar{q}$ resonance with a width over mass ratio of 0.10. This is a conservative choice since, as we also show in fig. \ref{fig:results}, the $G^{*}$ width over mass ratio is above 0.15 in most of the $G^{*}$ parameter space relevant for our study. \footnote{
More in details, we calculate the $G^{*}$ excluded regions by rescaling the $\sigma \times BR \times A$ values for Colorons/Axigluons ($C$) \cite{Simmons:1996fz} shown in \cite{Khachatryan:2015sja} according to 
\[
0.86 \times\frac{\sigma(G^{*})}{\sigma(C)}\frac{BR(G^{*} \to q\bar{q})}{(C\to q\bar{q})}= 0.86 \times\tan^2\theta_{3}\times \frac{BR(G^{*} \to q\bar{q})}{2/3}
\]
the universal Coloron model considered by CMS \cite{Simmons:1996fz} is indeed reproduced by the $G^{*}$ model in the limit: $\tan\theta_{3}=1$, $s_L=s_R=s_{bR}=0$ and decoupled top partners. The factor 0.86 is included to take into account the limitations of the narrow width approximation, as suggested in \cite{Khachatryan:2015sja}, and is taken from table 7 in \cite{Khachatryan:2015sja}. The resulting values are then compared to the upper limits for $q\bar{q}$ resonances with $\Gamma/M =0.1$.}
As expected from the $\tan^2\theta_{3}$ scaling behavior of the 
$G^{*}$ production cross section, we find that the searches for dijet resonances exclude only a part of the $G^{*}$ parameter space at large ($\gtrsim 0.5$) $\tan\theta_{3}$ values.  
The region of the $G^{*}$ parameter space excluded by dijet searches is colored grey in our final plot of fig. \ref{fig:results}.

\section{Applying the ATLAS selection}\label{sec:atlas}

In this section we apply the main requirements of the ATLAS selection in \cite{Aad:2015wqa} to Monte Carlo generated samples of the process $pp\to G^{*} \to B_Z \bar{B}_Z \to ZZ b\bar{b}$ in fig. \ref{fig:signal}. We generate the events for the heavy gluon signal at leading order with MADGRAPH 5 \cite{Alwall:2011uj}, after having implemented the model in sec. \ref{sec:model} by using feynrules \cite{Christensen:2008py}. We use the cetq6l1 PDF set \cite{Nadolsky:2008zw} with renormalization and factorization scales fixed at the heavy gluon mass. The events are then passed to PYTHIA 6.4 \cite{Sjostrand:2006za} (with the default tune) for showering and hadronization. Jets are reconstructed with FASTJET \cite{Cacciari:2011ma} by an anti-kt algorithm with cone size $R=$0.4. In order to mimic detector effects we also apply a Gaussian smearing to the jet energy with:
\begin{equation}
\frac{\sigma(E)}{E} = C + \frac{N}{E} + \frac{S}{\sqrt{E}}
\end{equation}
where $E$ is in GeV and $C=0.025$, $N=1.7$, $S=0.58$ \cite{Kulchitsky:2000gg}. The jet momentum is then rescaled by a factor $E^{smeared}/E$.\\

Retracing the ATLAS analysis, we consider the following final state: two same flavor (electrons or muons) leptons with opposite charge plus at least two jets and missing energy:

\begin{equation}
e^{+}e^{-}/ \mu^{+}\mu^{-} + n_{jet}\ \text{jets} + E^{miss}_T \qquad , \  n_{jet} \geq 2 \ .
\label{eq:final-state}
\end{equation}
\noindent
As a first step of the selection we apply the following isolation criteria and $p_T$ requirements on the two final leading leptons (we apply the same cuts for electrons and muons):

 \begin{align}
 \begin{split}
& |\eta (l_{1,2}) |<2.4 \qquad \Delta R (l_{1,2}, \text{jet})>0.3 \qquad \Delta R (l_1 ,l_2) > 0.3 \\
&\qquad p_T\ l_1 > 25 \ \text{GeV} \qquad p_T \ l_2 > 14 \ \text{GeV} \ .
\end{split}
\label{eq:acc-lep}
\end{align}
The jet separation requirement is applied to any ``baseline" jet with $p_T>$ 20 GeV and $|\eta|<5$. $l_1$ ($l_2$) denotes the leading (sub-leading) lepton.  \\
Signal jets must fulfill the conditions:
\begin{align}
 \begin{split}
& |\eta (\text{jet}) |<2.5 \qquad  p_T \ \text{jet} > 35 \ \text{GeV} \ .
\end{split}
\label{eq:acc-jet}
\end{align}

\begin{figure}
\centering
\includegraphics[width=0.4\textwidth]{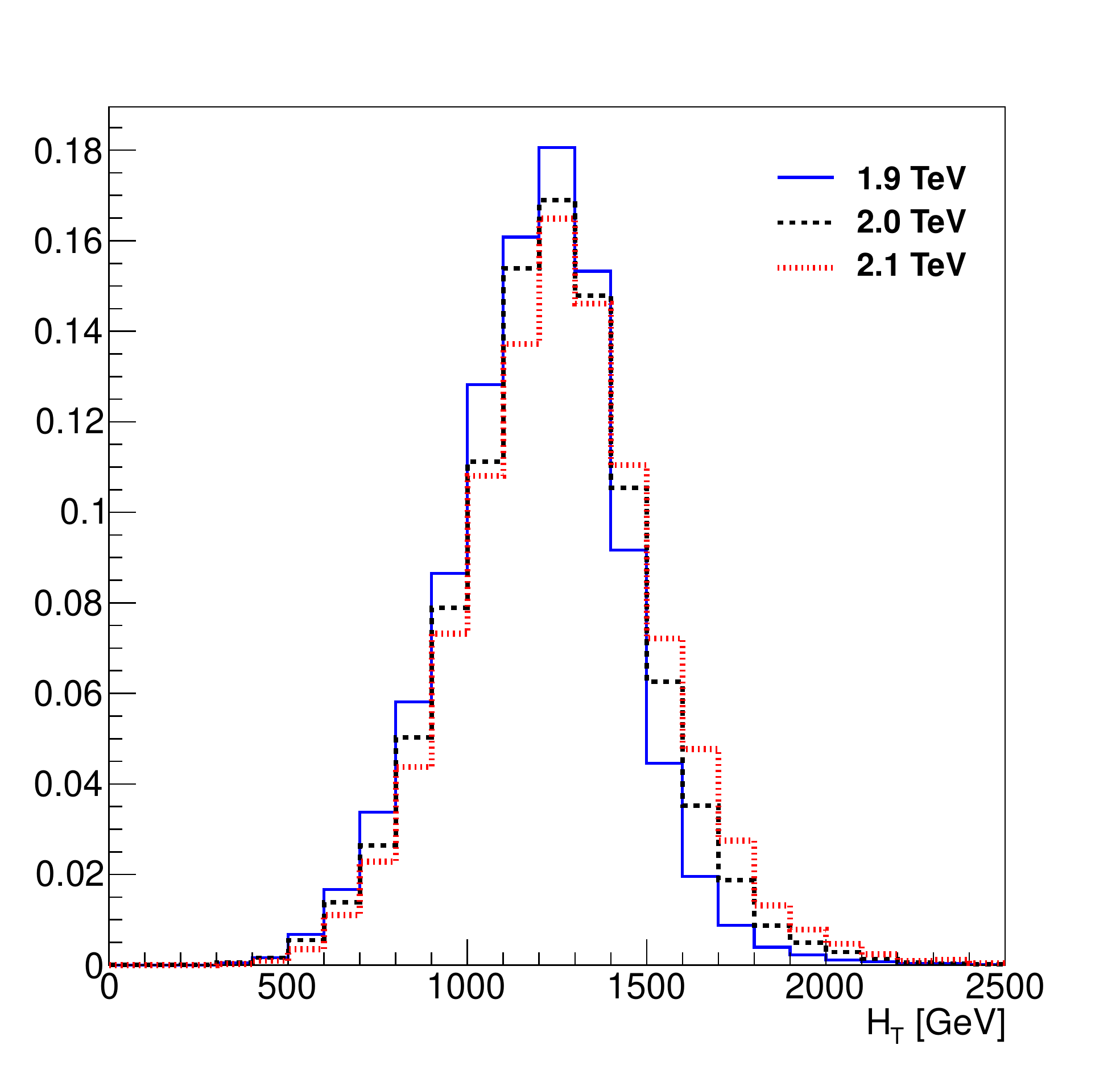} \includegraphics[width=0.4\textwidth]{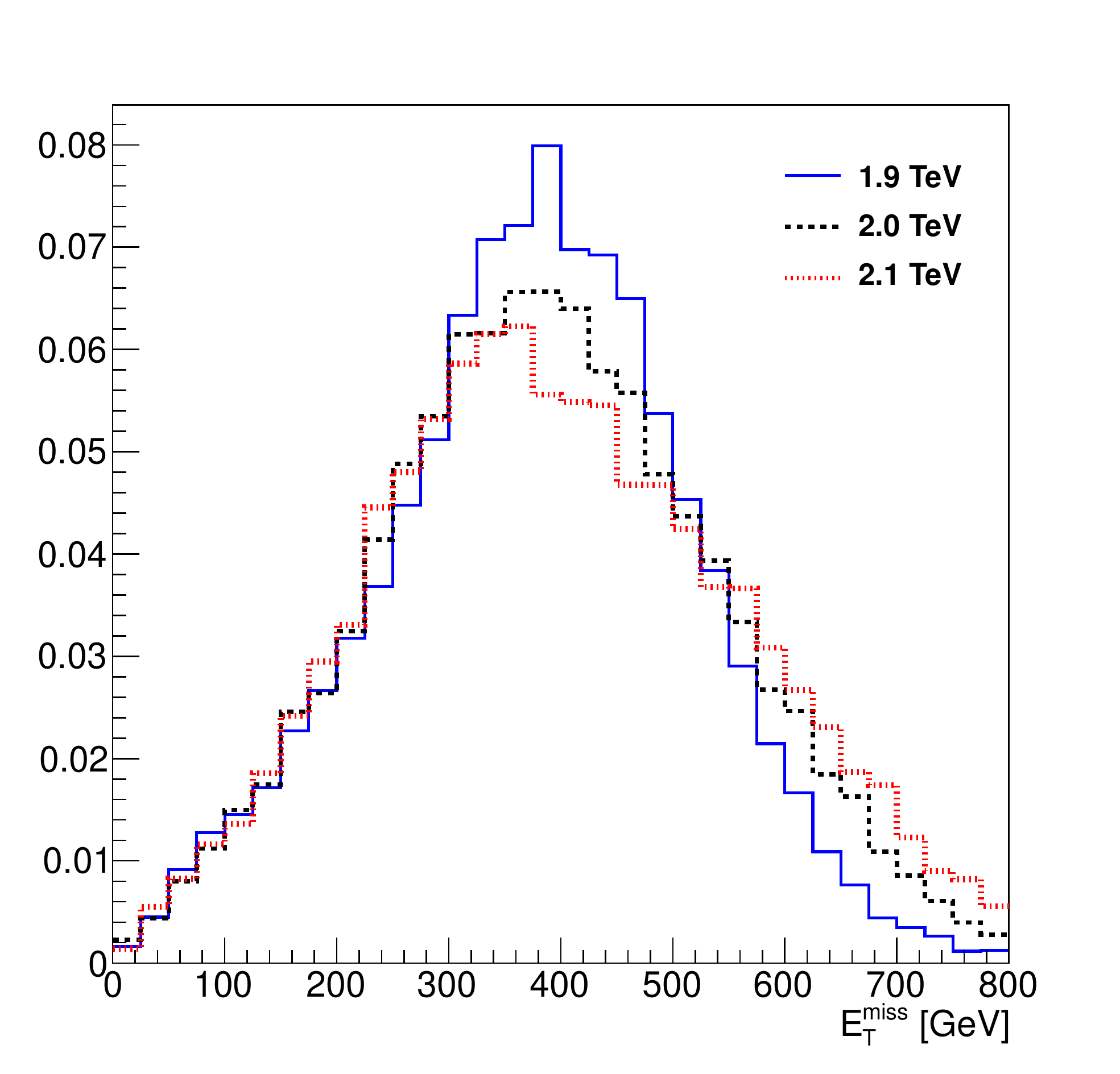}
\caption{\small Normalized $H_T$ (left plot) and $E^{miss}_T$ (right plot) distributions of $G^{*} \to B_Z\bar{B}_Z$ events in the same flavor opposite charge dilepton plus jets plus missing energy channel, eq. (\ref{eq:final-state}), after the acceptance cuts of eq. (\ref{eq:acc-lep}), (\ref{eq:acc-jet}). Distributions are shown for $G^{*}$ masses of 1.9, 2.0, 2.1 TeV and $\tan\theta_{3}=0.5$. }
\label{fig:HT-ETmiss-norm}
\end{figure}

After these ``acceptance" cuts, we get, at $\sqrt{s}=8$ TeV LHC with 20.3 fb$^{-1}$, the number of $G^{*}$ signal events shown on the second column of table \ref{tab:cut-flow}. The expected number of events is shown for several heavy gluon masses and for a fixed coupling $\tan\theta_{3}=0.5$.\\
The on-Z ATLAS search is focused on generalized gauge mediated supersymmetry breaking signatures, where pair-produced gluinos decay via neutralinos to a gravitino, which is the lightest supersymmetry particle, plus a Z boson. These GGM topologies are characterized by energetic final states with large missing energy associated with the gravitino.  
The ATLAS analysis thus applies the following set of cuts, aiming at selecting events with Z-peaked dilepton mass, large missing energy and energetic final states:

\begin{align}
 \begin{split}
& 81 \ \text{GeV} < m_{ll} < 101 \ \text{GeV} \qquad E^{miss}_T  > 225\ \text{GeV} \\
&  \qquad H_T  > 600 \ \text{GeV} \qquad \Delta\phi(\text{jet}_{1,2}, E^{miss}_T) > 0.4 \ ,
\end{split}
\label{eq:cuts}
\end{align}
where $H_T$ is defined as: $H_T = p_T (l_1 ) + p_T (l_2 ) + \sum^{n_{jet}}_{i} p_T (\text{jet}^{i}) $
, $m_{ll}$ is the dilepton mass, $E^{miss}_T$ denotes the transverse missing energy and $\Delta\phi(\text{jet}_{1,2}, E^{miss}_T) $ is the azimuthal opening angle between the missing energy and the leading or sub-leading jet; the restriction on the azimuthal jet-$E^{miss}_T$ separation is applied to reject events with jet mismeasurements
contributing to large fake missing energy. \\
We find that the ATLAS search strategy, with the main set of cuts in (\ref{eq:cuts}), does not only apply to GGM topologies but also selects heavy gluon $G^{*} \to B_Z \bar{B}_Z \to ZZ b\bar{b}$ signals (fig. \ref{fig:signal}), which are also characterized by a leptonically decaying $Z$, an energetic final state and large missing energy, produced by the decay to neutrinos of one of the two $Z$ bosons in the final state. The $H_T$ and $E^{miss}_T$ distributions for the $G^{*} \to B_Z\bar{B}_Z$ events in the channel (\ref{eq:final-state}) after the acceptance cuts in (\ref{eq:acc-lep}), (\ref{eq:acc-jet}) are shown in fig. \ref{fig:HT-ETmiss-norm} for several $G^{*}$ masses \footnote{We calculate $E^{miss}_T$ by Monte Carlo truth: we sum vectorially the transverse momentum of the neutrinos and those of soft ($p_T<$ 20 GeV) or lost ($|\eta|>$5) jets.}. \\
Applying the ATLAS selection (\ref{eq:cuts}) to the $G^{*}$ signal, we find the expected number of events, at the 8 TeV LHC with 20.3 fb$^{-1}$, indicated in tab. \ref{tab:cut-flow}. Columns 3-6 show the number of events passing the different steps of the ATLAS selection for several $G^{*}$ masses at a fixed coupling $\tan\theta_{3}=0.5$.\\

\begin{table}[t]
\begin{center}
{\small
\begin{tabular}[]{c|ccccc|c}
$M_{G^{*}}$ (GeV) & $n_{jet}$ & $m_{ll}$ & $E^{miss}_T$ & $H_T$ & $\Delta\phi(\text{jet}_{1,2}, E^{miss}_T)$ & $\tan\theta_{3}$\\[0.1cm]
[$\tan\theta_{3}=0.5$] & $\geq$ 2 & [81 , 101] & $>$225 & $>$600 & $>$0.4 & [$n^{exc}$= 18.4 $\pm$ 3.2] \\
\hline  
&&&&&&  \\[-0.3cm]
1870 & 68 & 60 & 52 & 51 & 46 & 0.29 $\pm$ 0.03 \\[0.1cm]
1900 & 56 & 49 & 42 & 42 & 38 & 0.33 $\pm$ 0.03 \\[0.1cm]
1950 & 38 & 33 & 28 & 28 & 25 & 0.41 $\pm$ 0.04 \\[0.1cm]
2000 & 30 & 26 & 23 & 23 & 20 & 0.47 $\pm$ 0.05 \\[0.1cm]
2100 & 21 & 19 & 16 & 16 & 14 & 0.58 $\pm$ 0.07 \\[0.1cm]
2150 & 18 & 16 & 13 & 13 & 12 & 0.65 $\pm$ 0.09 \\[0.1cm]
2200 & 16 & 14 & 12 & 12 & 11 & 0.74 $\pm$ 0.14 
\end{tabular}
}
\caption{ \small Number of events at the 8 TeV LHC with 20.3 fb$^{-1}$ for the $G^{*} \to B_Z \bar{B}_Z$ signal in the same flavor opposite charge dilepton channel which pass the different steps of the ATLAS selection \cite{Aad:2015wqa} ($m_{ll}$, $E^{miss}_{T}$ and $H_{T}$ cuts are in GeV). We show the results for several $G^{*}$ masses at a fixed coupling $\tan\theta_{3}=0.5$. The last column indicates, for each $G^{*}$ mass point, the $\tan\theta_{3}$ value which gives the excess of events measured by ATLAS \cite{Aad:2015wqa}.
\label{tab:cut-flow}}
\end{center}
\end{table}

\section{Results}\label{sec:results}

After the complete selection in (\ref{eq:cuts}) with 20.3 fb$^{-1}$ of collected integrated luminosity, 
ATLAS observes an excess of events above the expected Standard Model background with a statistical significance of 3 standard deviations. The statistical significance is of 3 sigma in the electron channel and of 1.7 sigma in the muon channel. The excess of events above background in the $ee+\mu\mu$ channel, which can be read from tab. 7 in \cite{Aad:2015wqa}, is:
\begin{equation}
n^{exc}= 18.4 \pm 3.2 \ .
\label{eq:exc}
\end{equation}

\begin{figure}
\centering
\includegraphics[width=0.75\textwidth]{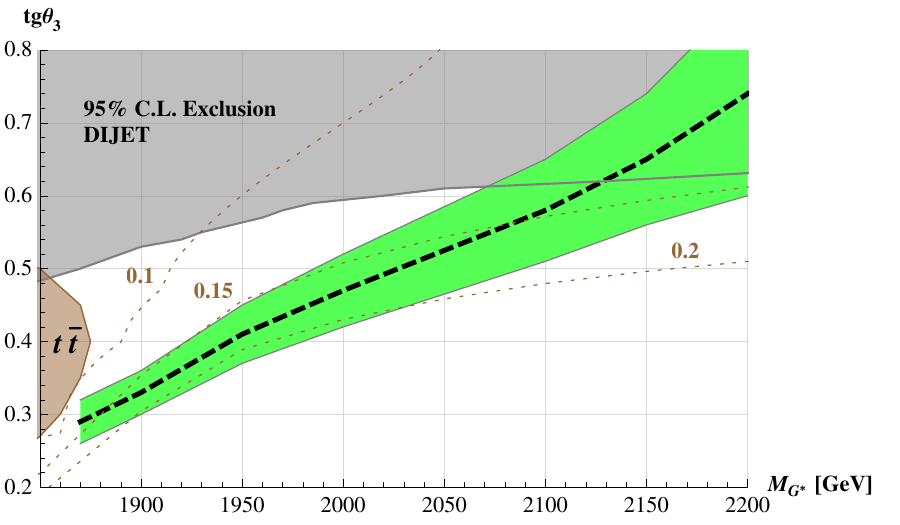}
\caption{\small $(M_{G^{*}}, \tan\theta_3)$ values reproducing the ATLAS \cite{Aad:2015wqa} excess of events. The green band shows the values giving the excess of events within $\pm 1 \sigma$ from the central value (black dashed curve). 
The grey upper region is excluded by searches for dijet resonances -- it is derived from the CMS limits on $q\bar{q}$ resonances with $\Gamma/M=0.1$ \cite{Khachatryan:2015sja}; the brown region on the left is excluded by searches for $t\bar{t}$ resonances \cite{CMS:2015nza}. Dotted curves show $(M_{G^{*}}, \tan\theta_3)$ regions 
 where $\Gamma(G^{*})/M_{G^{*}}$= 0.10, 0.15, 0.20. }
\label{fig:results}
\end{figure}

By applying the analysis explained in the previous section to $G^{*} \to B_Z \bar{B}_Z$ events for different heavy gluon masses and couplings we can derive the $(M_{G^{*}}, \tan\theta_{3})$ values which are able to reproduce the ATLAS results. The last column in tab. \ref{tab:cut-flow} indicates, for the different $G^{*}$ mass points, the $\tan\theta_{3}$ values which give the excess of events in (\ref{eq:exc}) measured by ATLAS. \\
Our results are finally shown in fig. \ref{fig:results}. The green band shows the $(M_{G^{*}}, \tan\theta_3)$ values giving the excess of events in (\ref{eq:exc}) within $\pm 1 \sigma$ from the central value, indicated by the black dashed curve. The grey upper region is excluded by searches for dijet resonances, calculated, as explained in sec. \ref{sec:excl}, from the CMS search \cite{Khachatryan:2015sja}. The lower brown region is excluded by searches for $t\bar{t}$ resonances and is derived (sec. \ref{sec:excl}) from the CMS analysis in\cite{CMS:2015nza}. We see that heavy gluon decays to vector-like bottom quarks can explain the ATLAS finding in a parameter space region so far untested and thus not-excluded by LHC searches. We find that the ATLAS excess of events can be produced by $G^{*} \to B_Z \bar{B}_Z$ events for $G^{*}$ masses roughly in a range 1.87 - 2.15 TeV and for $g_S \tan\theta_3$ couplings within $\sim(0.3-0.65) g_S$. \\

\begin{figure}
\centering
\includegraphics[width=0.47\textwidth]{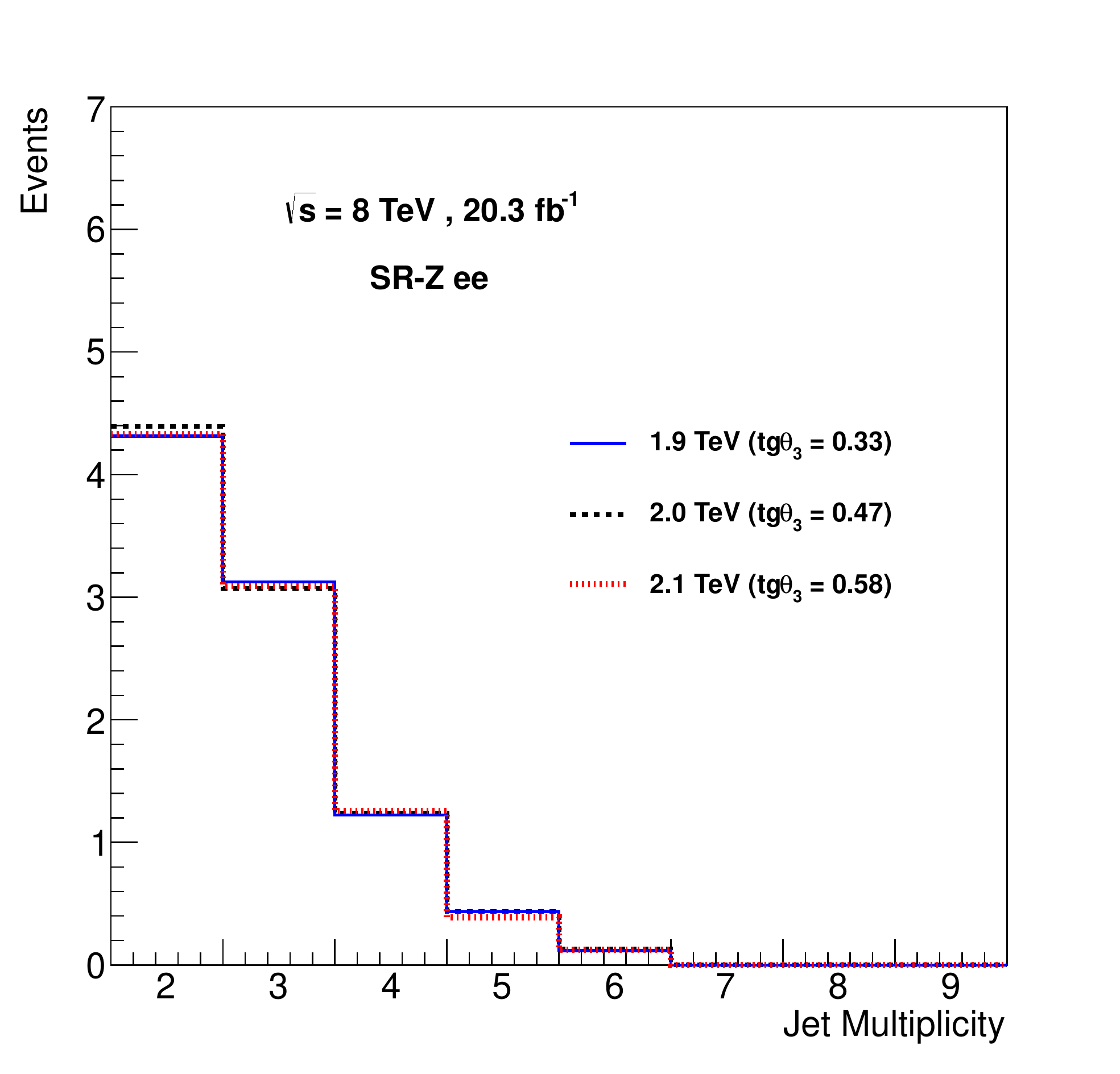}
\includegraphics[width=0.45\textwidth]{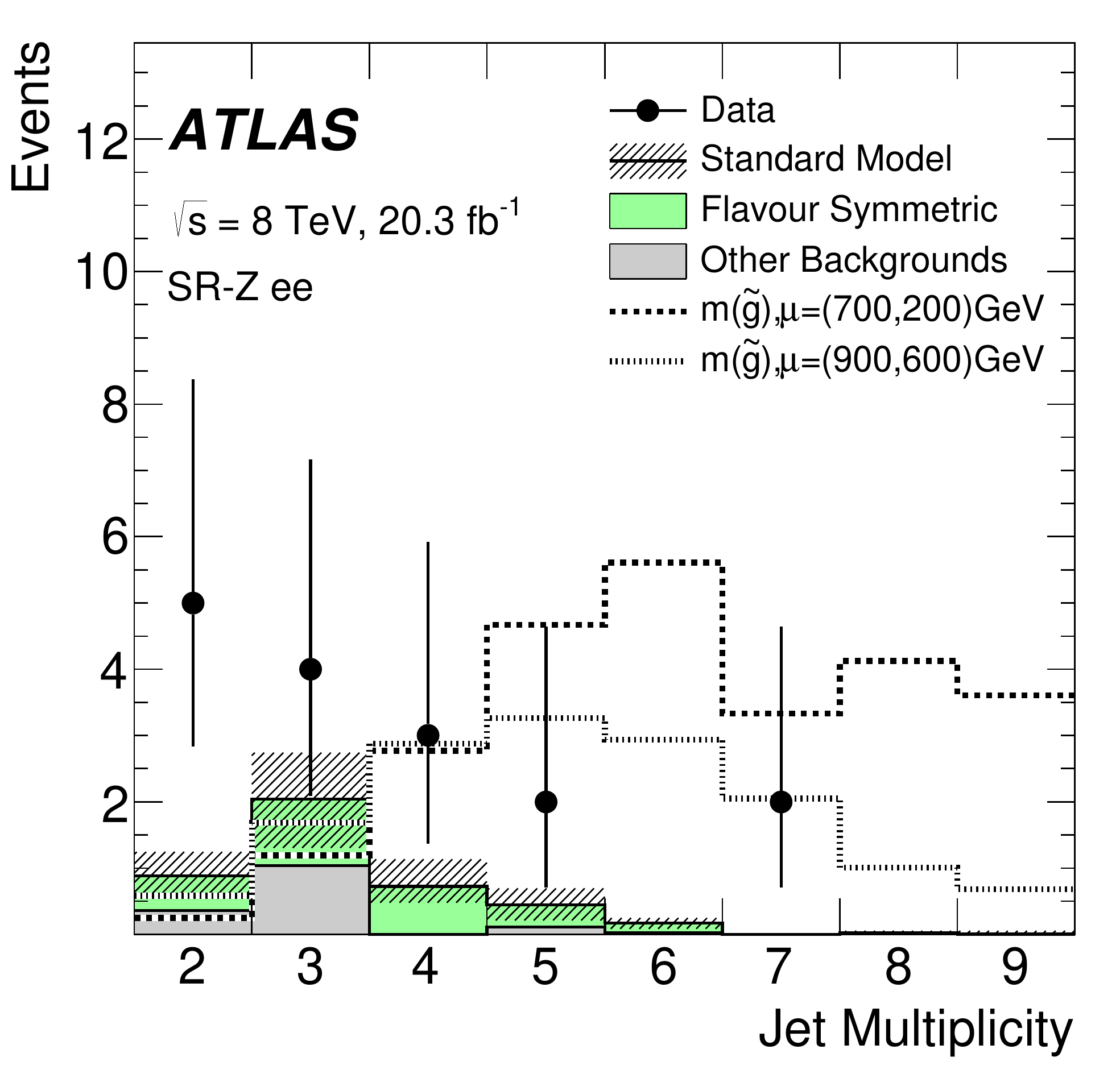}
\includegraphics[width=0.47\textwidth]{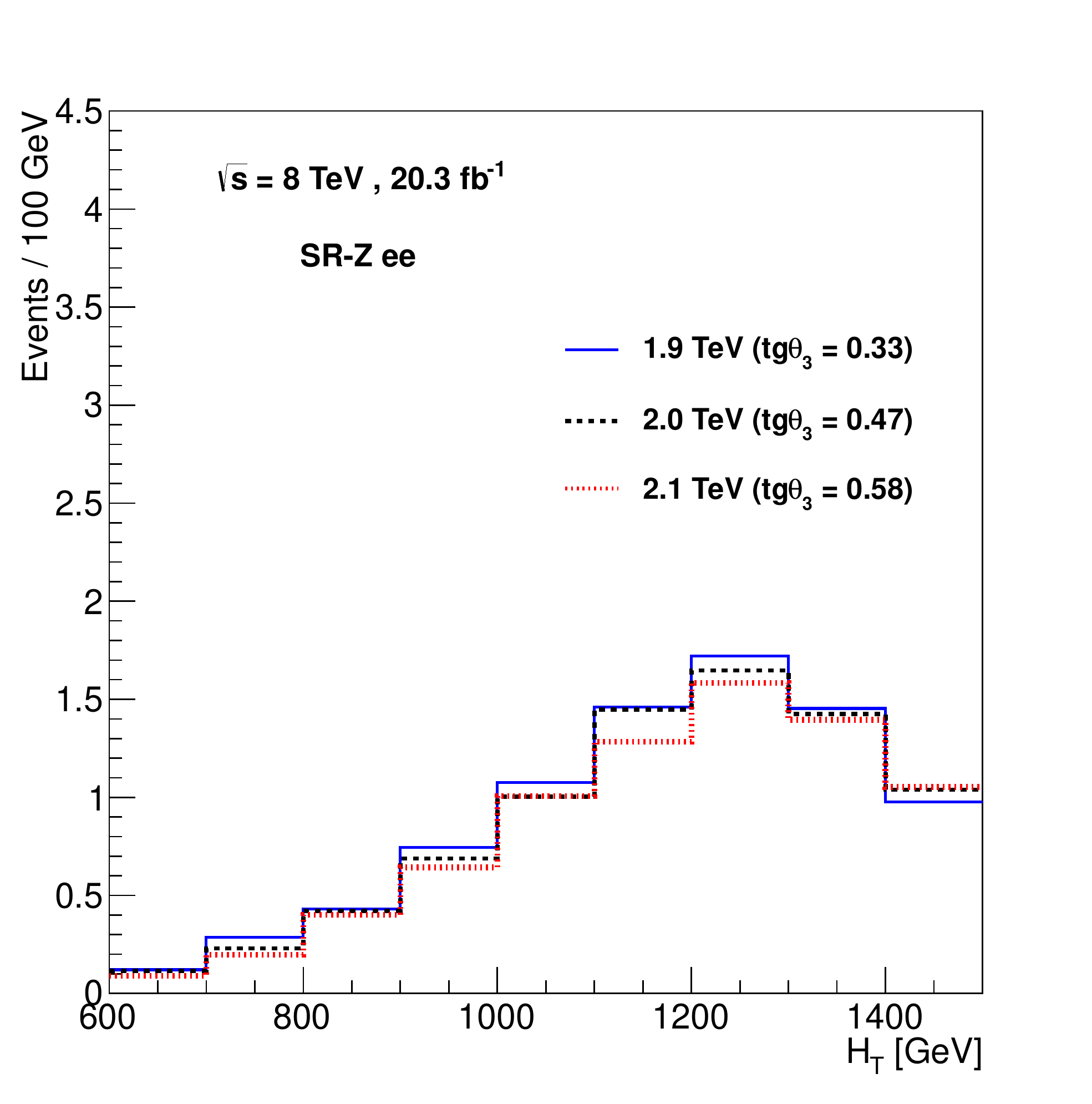}
\includegraphics[width=0.45\textwidth]{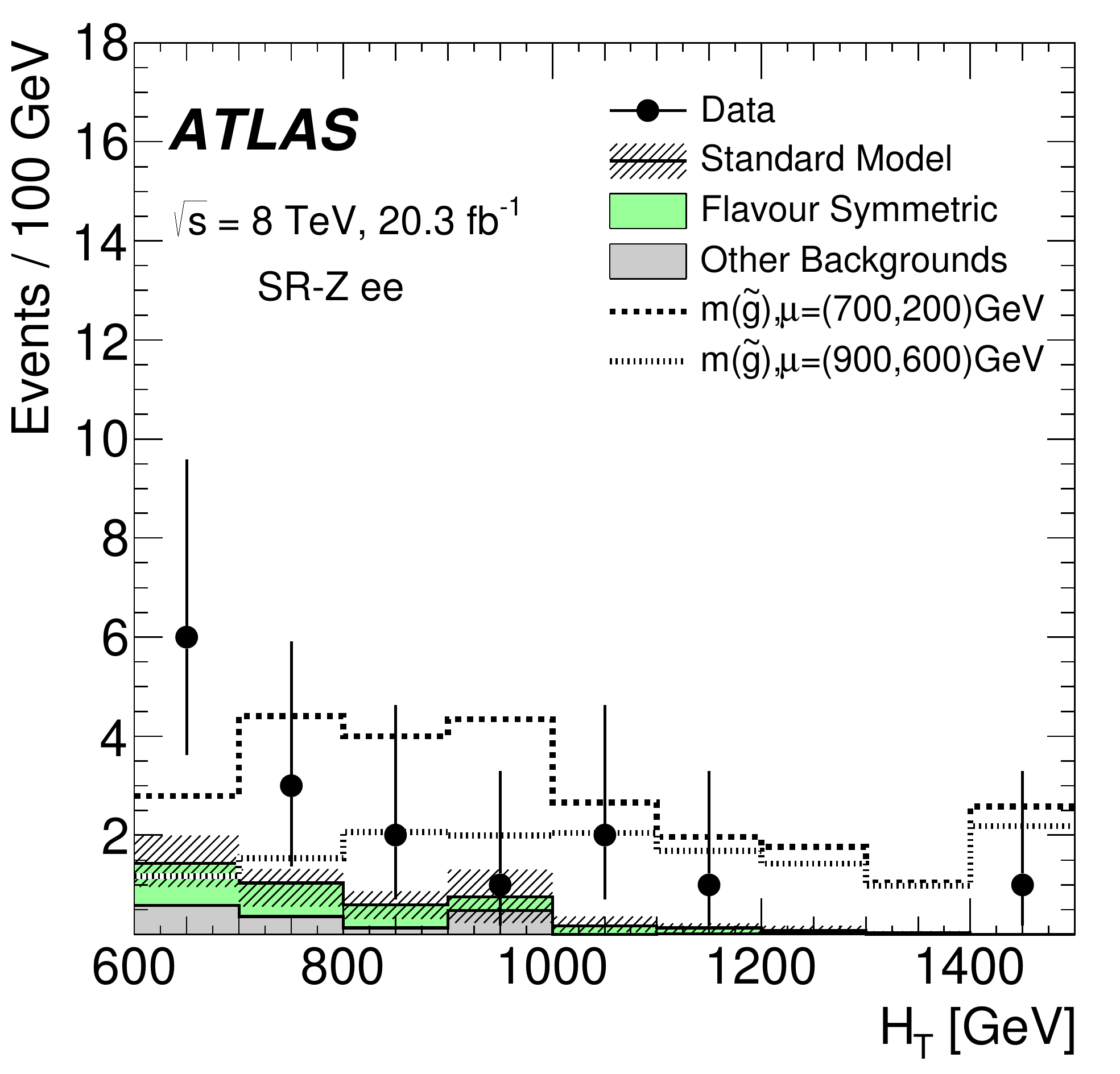}
\caption{\small Kinematic distributions of events (only for the $ee$ channel): jet multiplicity in the upper plots and $H_T$ in the lower plots. The distributions are obtained after all of the cuts in (\ref{eq:cuts}) have been applied. The Plots on the left refer to the $pp\to G^{*} \to B_Z \bar{B}_Z \to ZZ b\bar{b}$ signal for different heavy gluon masses (1.9 TeV, 2.0 TeV, 2.1 TeV) and for the $\tan\theta_3$ values reproducing (the central value of) the excess of events measured by ATLAS \cite{Aad:2015wqa} in the $ee+\mu\mu$ channel. The Plots on the right are taken from ATLAS \cite{Aad:2015wqa} and show the jet multiplicity and $H_T$ distributions of the observed events, of the estimated backgrounds and of two examples of GGM models with $\tan\beta$= 1.5.  } 
\label{fig:dist1}
\end{figure}

\begin{figure}
\centering
\includegraphics[width=0.48\textwidth]{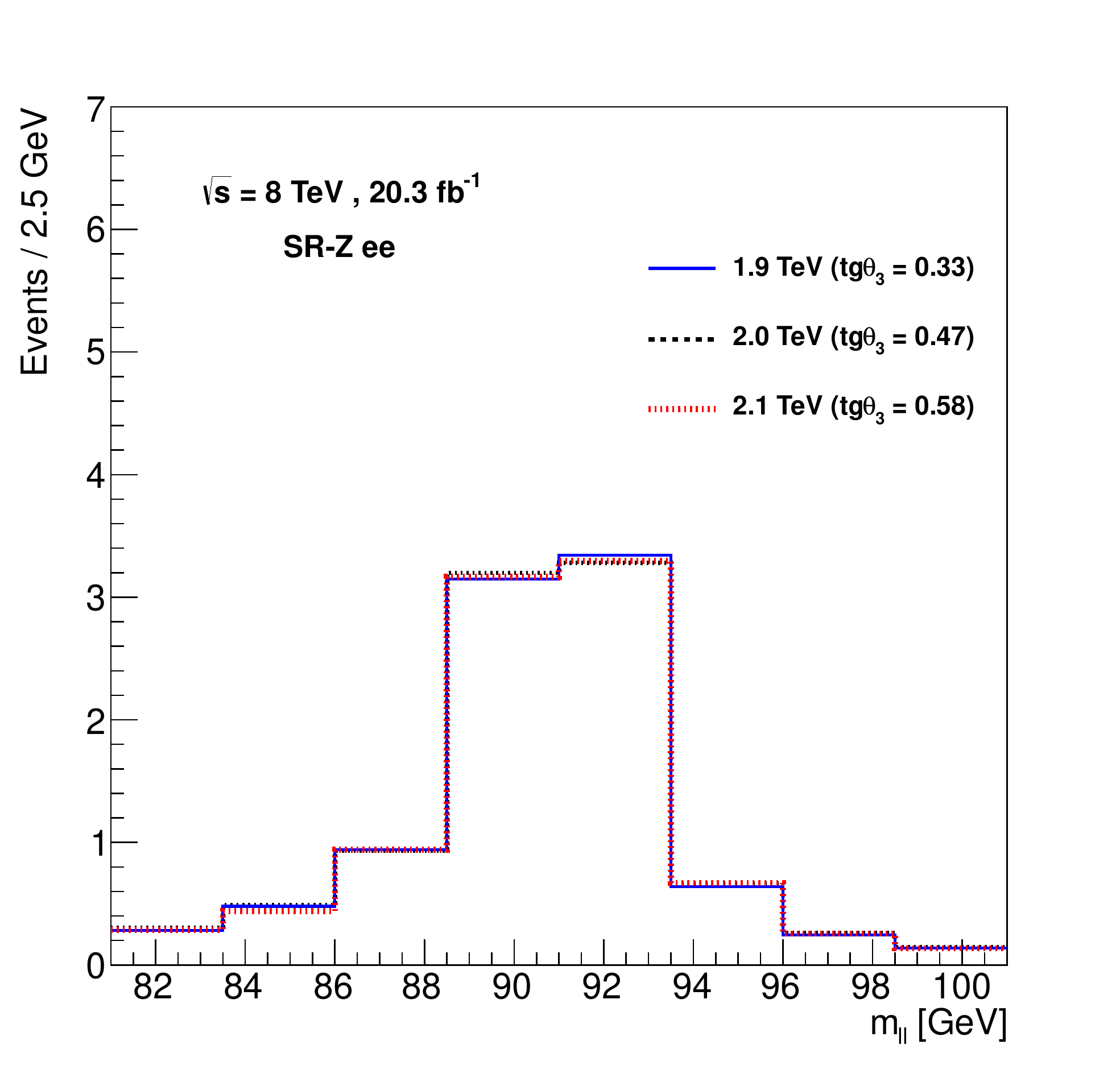}
\includegraphics[width=0.45\textwidth]{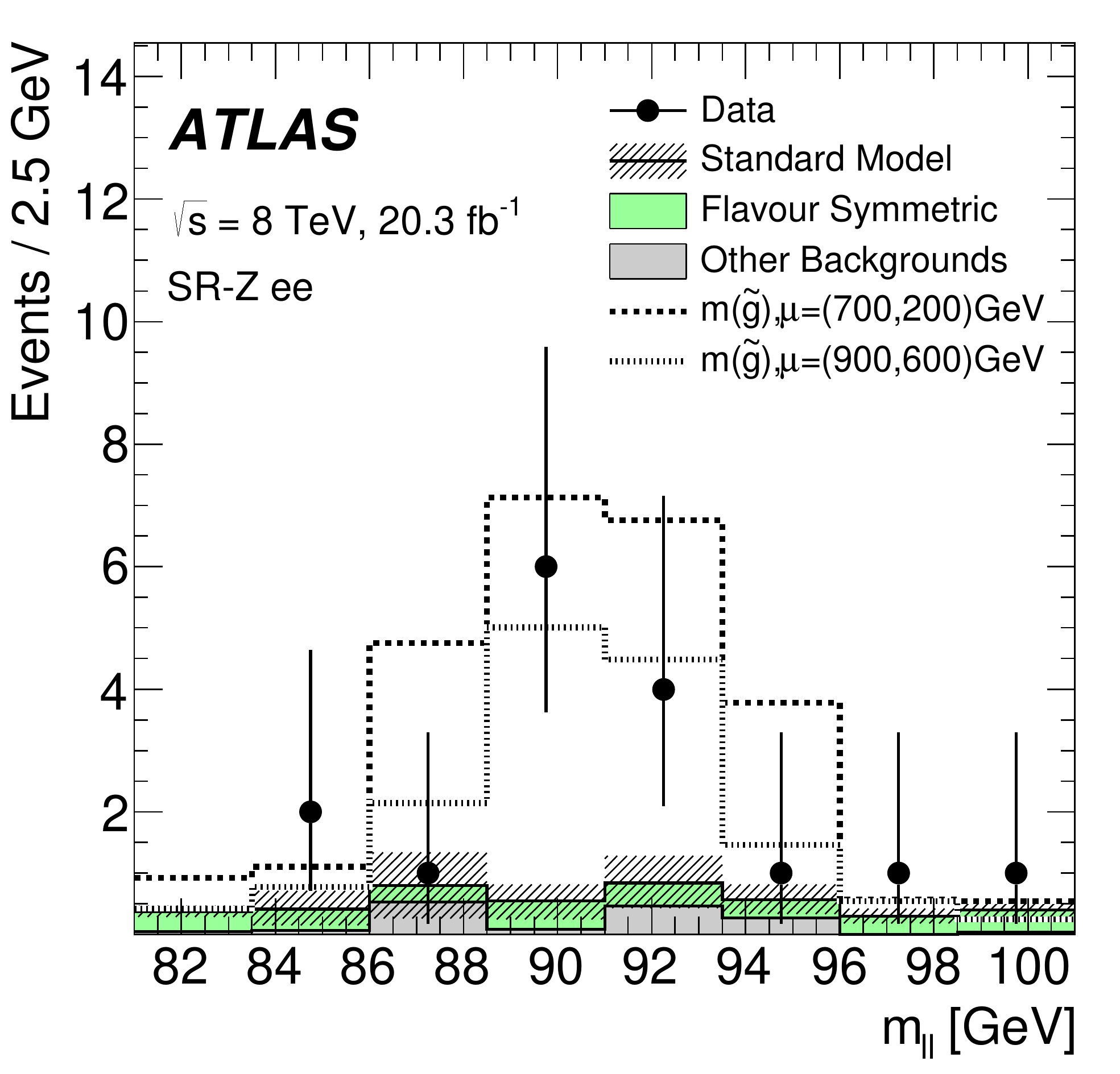}
\includegraphics[width=0.47\textwidth]{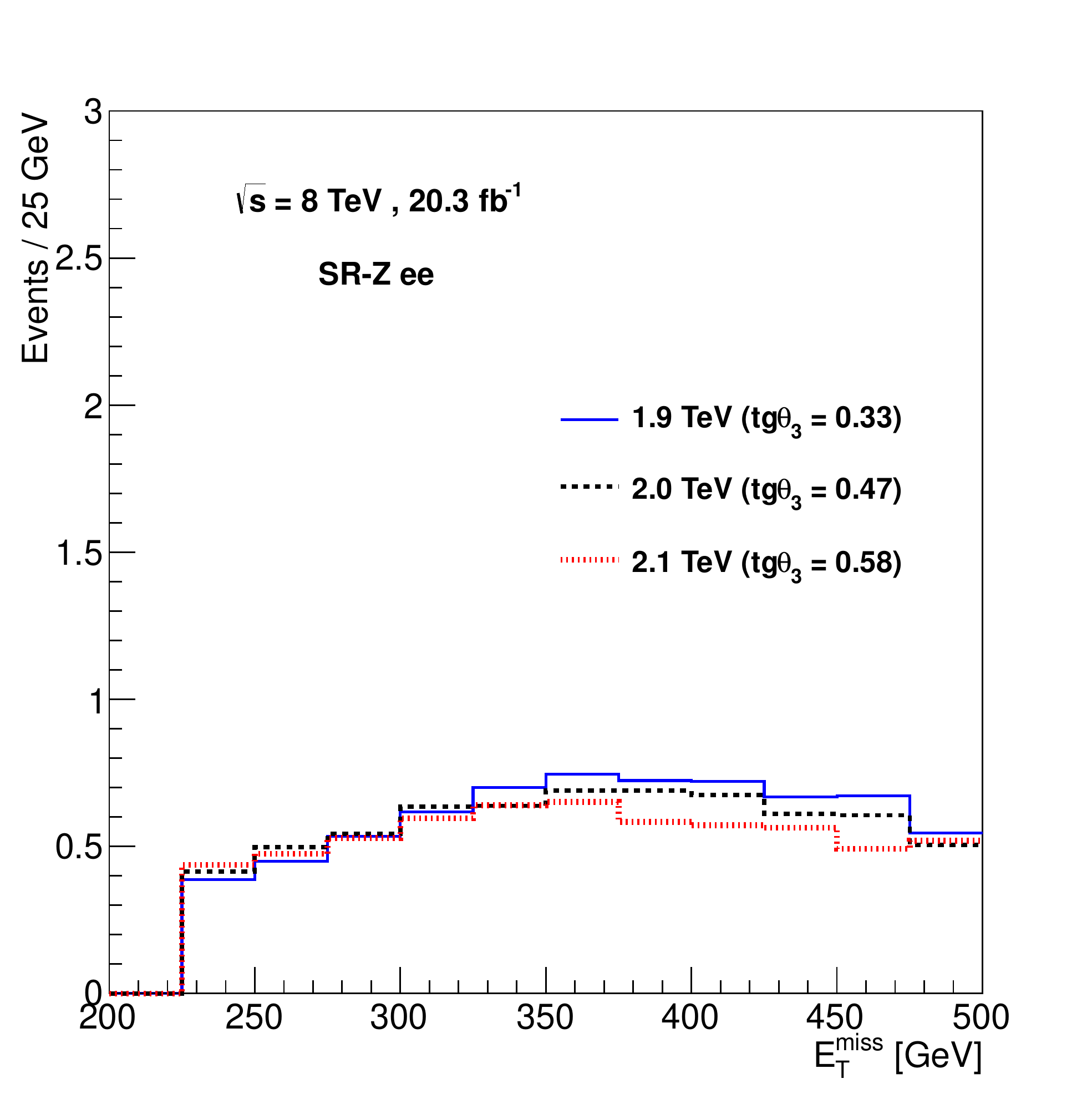}
\includegraphics[width=0.45\textwidth]{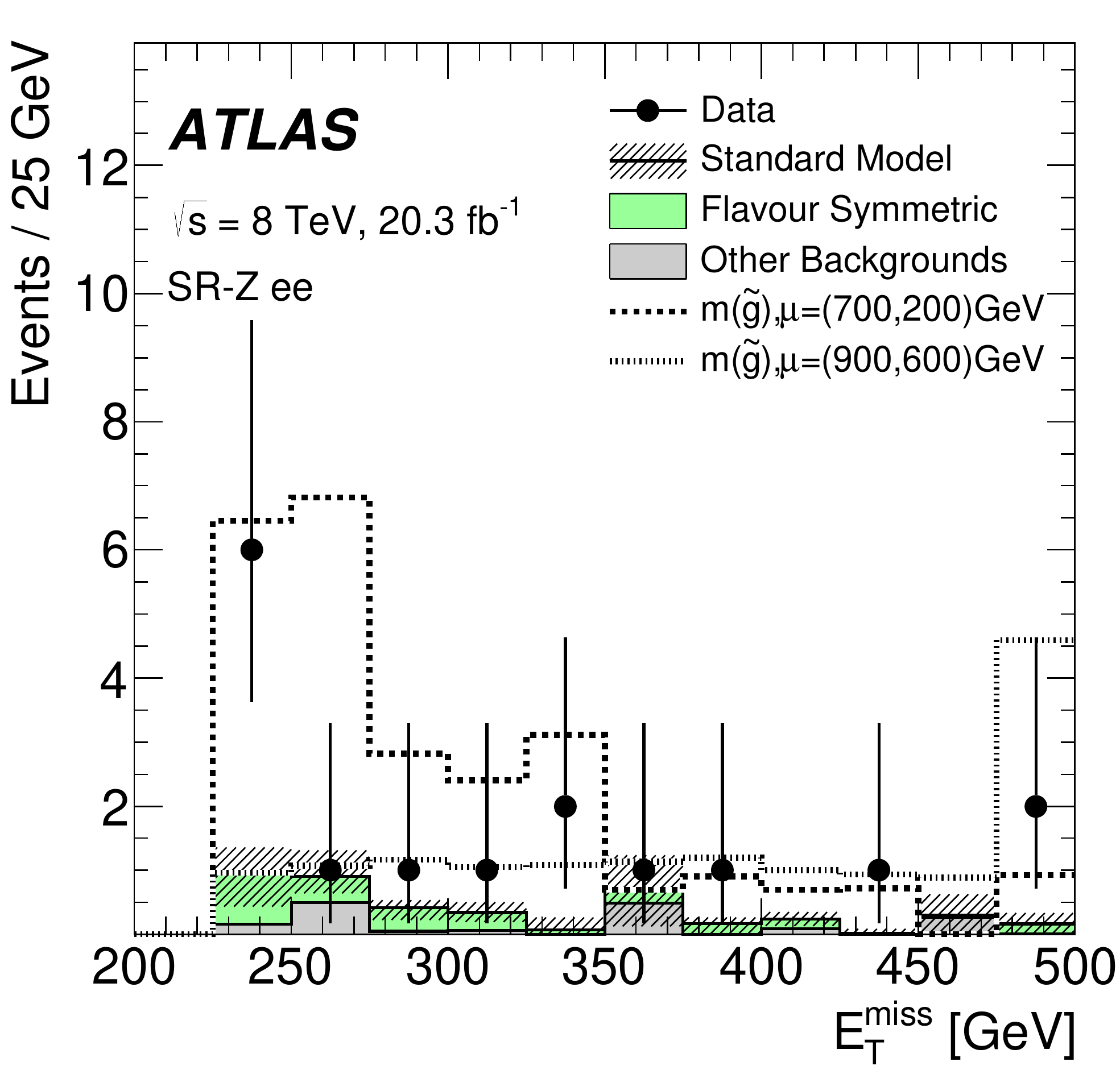}
\caption{\small Kinematic distributions of events (only for the $ee$ channel): dilepton mass in the upper plots and missing transverse energy in the lower plots. The distributions are obtained after all of the cuts in (\ref{eq:cuts}) have been applied. The Plots on the left refer to the $pp\to G^{*} \to B_Z \bar{B}_Z \to ZZ b\bar{b}$ signal for different heavy gluon masses (1.9 TeV, 2.0 TeV, 2.1 TeV) and for the $\tan\theta_3$ values reproducing (the central value of) the excess of events measured by ATLAS \cite{Aad:2015wqa} in the $ee+\mu\mu$ channel. The Plots on the right are taken from ATLAS \cite{Aad:2015wqa} and show the $m_{ll}$ and $E^{miss}_T$ distributions of the observed events, of the estimated backgrounds and of two examples of GGM models with $\tan\beta$= 1.5.} 
\label{fig:dist2}
\end{figure}

\begin{figure}
\centering
\includegraphics[width=0.48\textwidth]{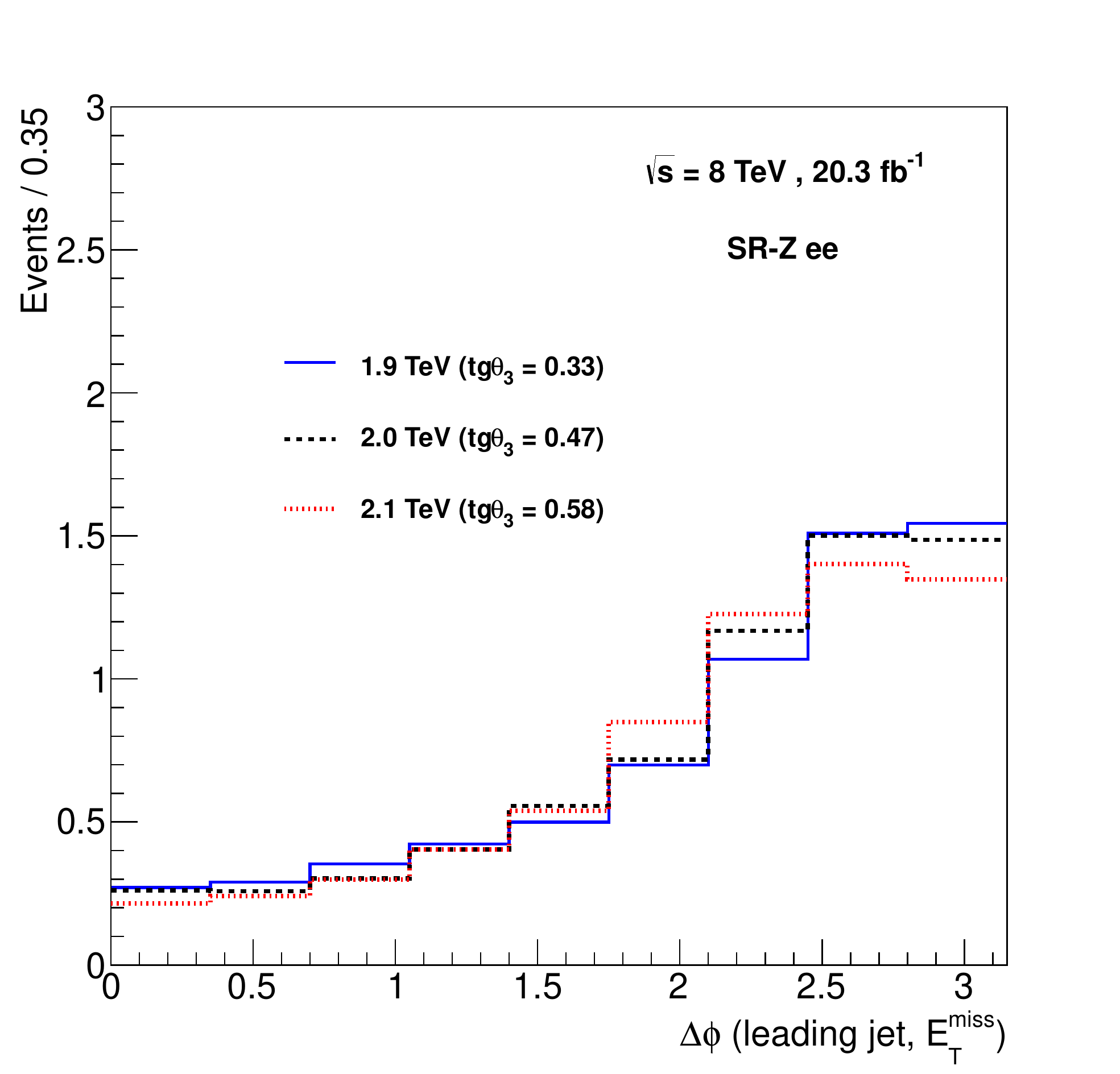}
\includegraphics[width=0.45\textwidth]{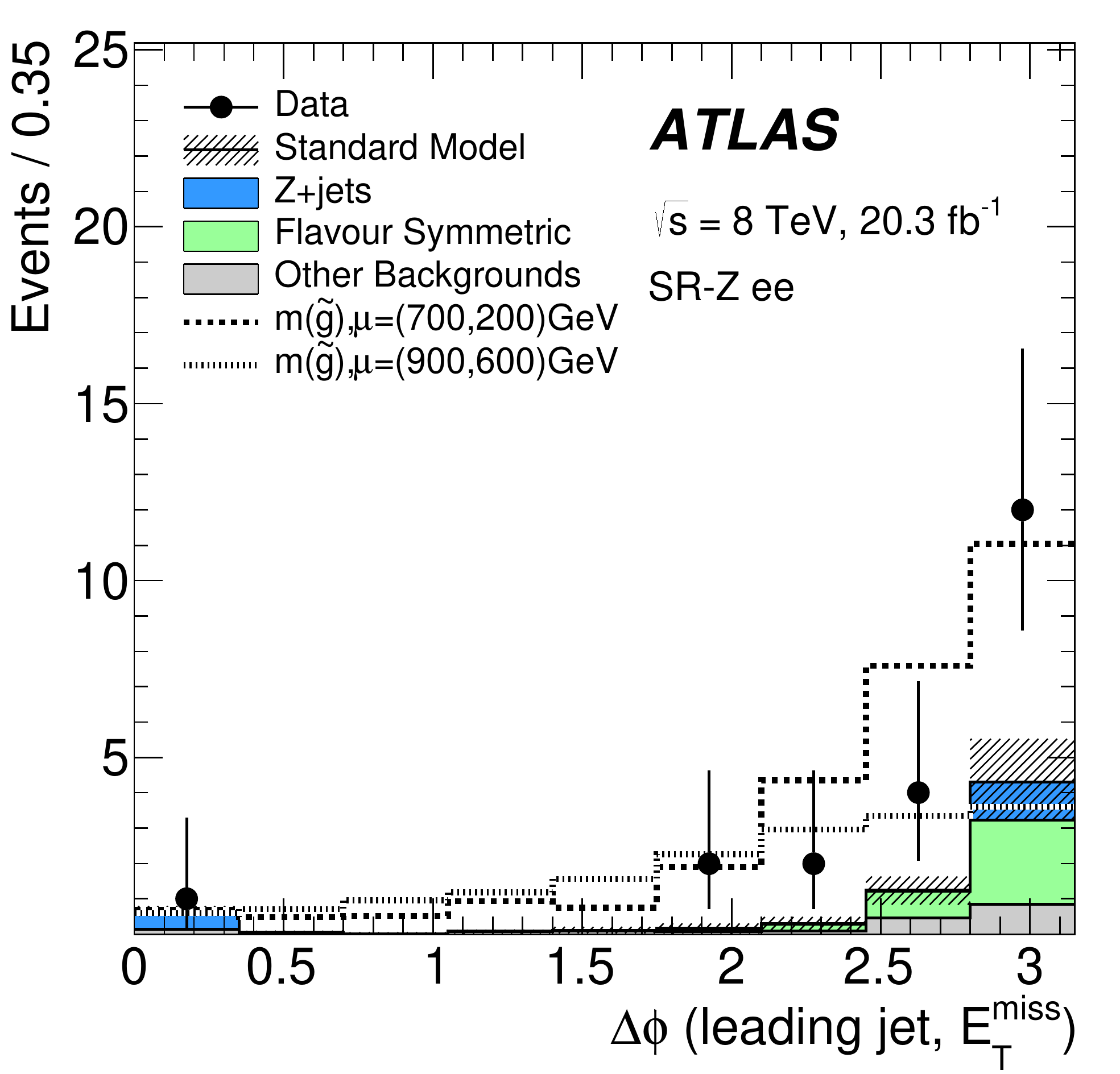}
\includegraphics[width=0.47\textwidth]{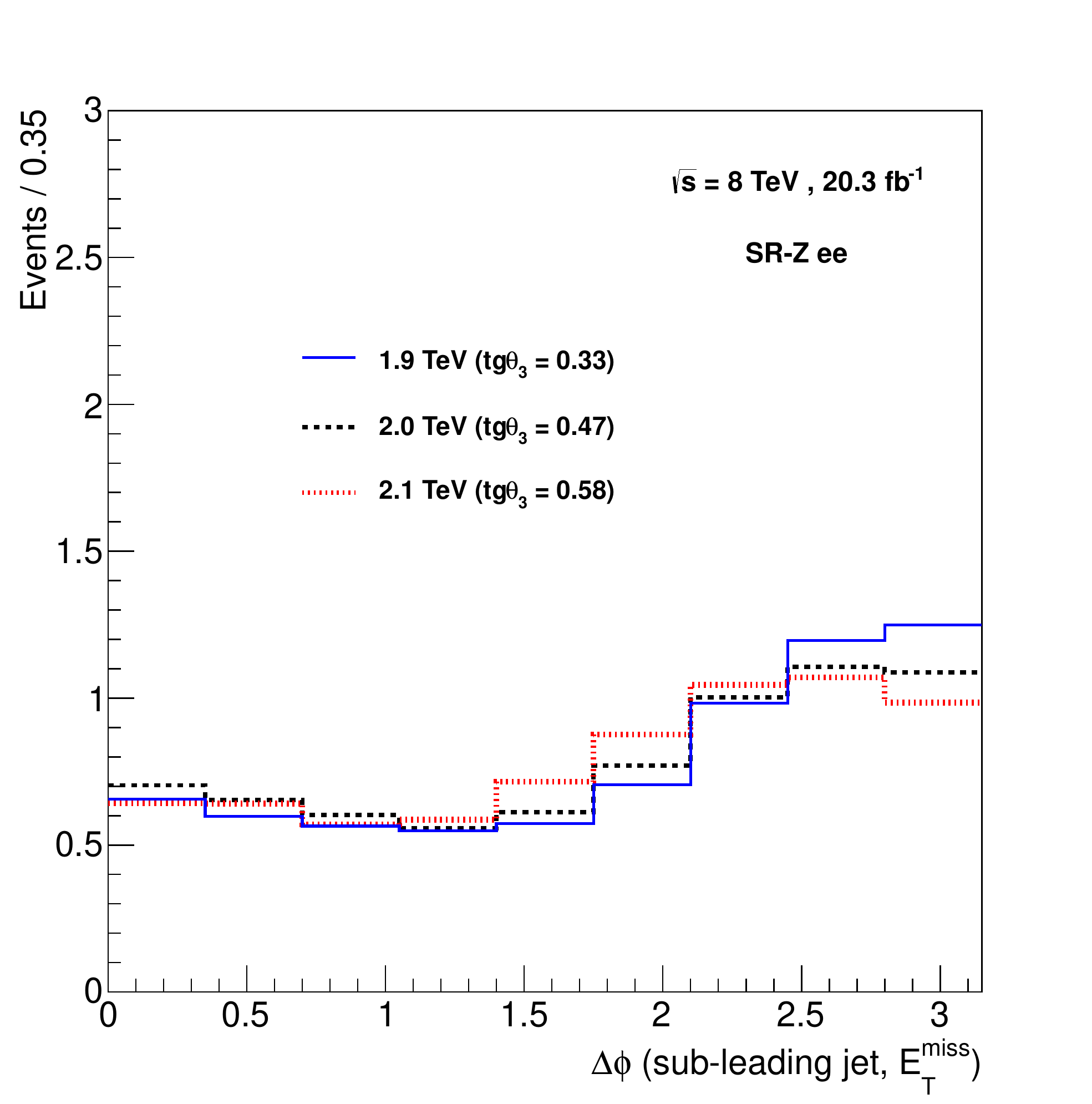}
\includegraphics[width=0.45\textwidth]{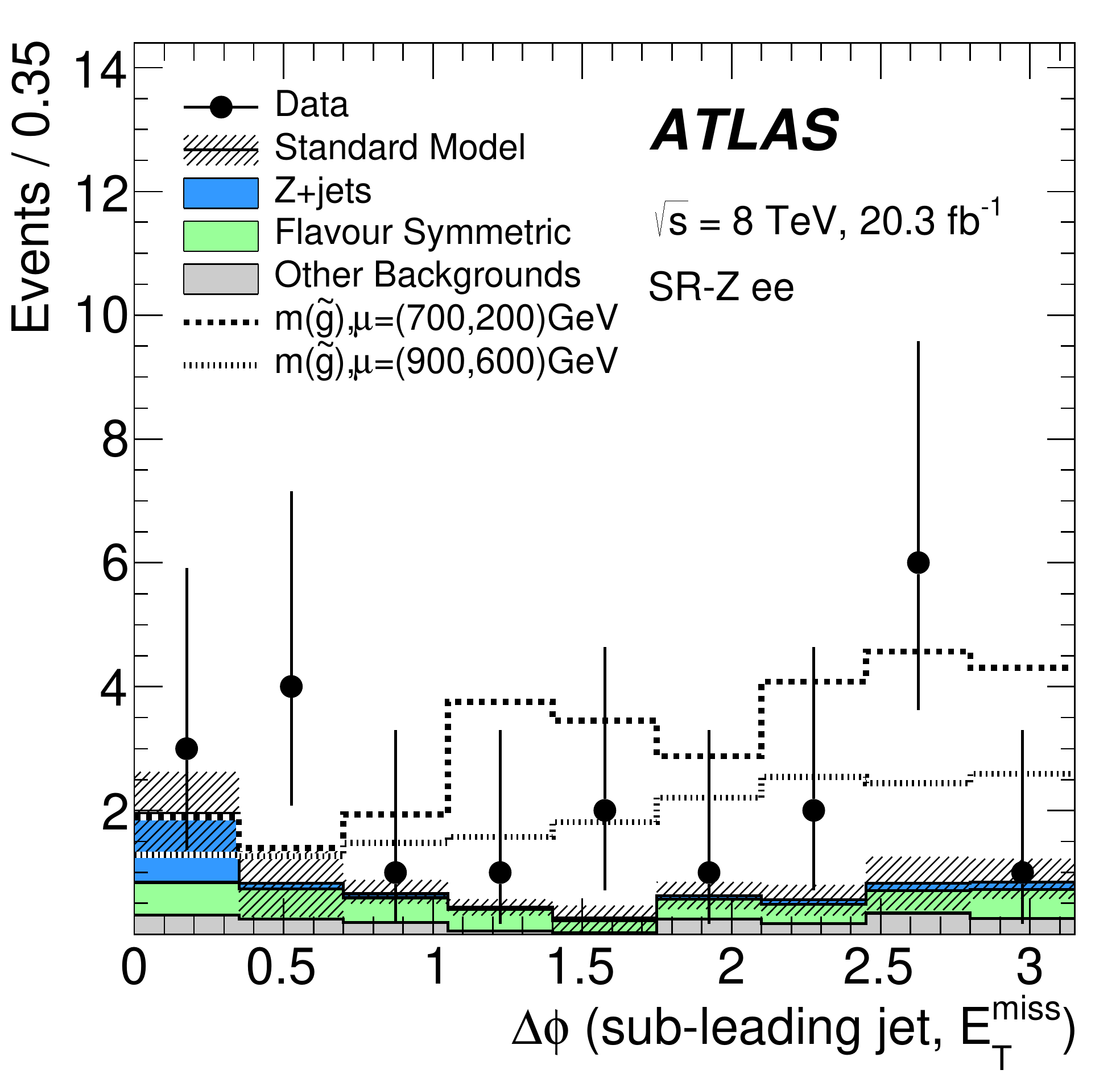}
\caption{\small Kinematic distributions of events (only for the $ee$ channel): azimuthal opening angle between the missing energy and the leading (upper plots) or sub-leading (lower plots) jet. The distributions are obtained after all of the cuts in (\ref{eq:cuts}), except the requirement $\Delta\Phi (\text{jet}_{1,2}, E^{miss}_T)>$0.4, have been applied. The Plots on the left refer to the $pp\to G^{*} \to B_Z \bar{B}_Z \to ZZ b\bar{b}$ signal for different heavy gluon masses (1.9 TeV, 2.0 TeV, 2.1 TeV) and for the $\tan\theta_3$ values reproducing (the central value of) the excess of events measured by ATLAS \cite{Aad:2015wqa} in the $ee+\mu\mu$ channel. The Plots on the right are taken from ATLAS \cite{Aad:2015wqa} and show the $\Delta\Phi (\text{jet}_{1,2}, E^{miss}_T)$ distributions of the observed events, of the estimated backgrounds and of two examples of GGM models with $\tan\beta$= 1.5} 
\label{fig:dist3}
\end{figure}

Finally, we show in fig.s \ref{fig:dist1},\ref{fig:dist2},\ref{fig:dist3} the kinematic distributions of $G^{*} \to B_Z \bar{B}_Z$ events, to be compared with the distributions of the observed events shown by ATLAS in \cite{Aad:2015wqa} -- which we report for the reader convenience on the right side of fig.s \ref{fig:dist1},\ref{fig:dist2},\ref{fig:dist3}. Since  the highest statistical significance is found in the electron channel, we just show the distribution in the $e^{+}e^{-}$ channel.
We find that $G^{*} \to B_Z \bar{B}_Z$ event distributions are compatible with 
the jet multiplicity, the $H_T$, the $E^{miss}_T$, the $m_{ll}$ and the $\Delta\phi(\text{jet}_{1,2}, E^{miss}_T)$ distributions of the ATLAS observed events \footnote{Obviously, $G^{*}$ signal events have to be summed to the expected background events in order to be compared with data.} with few exceptions: the number of observed events on the first bin in the $H_T$ distribution and on the first bin in the $E^{miss}_T$ distribution are slightly above the expected number of $G^{*} \to B_Z \bar{B}_Z$ plus background events \footnote{Here we just comment qualitatively on our findings for the event distributions. A detailed analysis of the statistical compatibility between the $G^{*} \to B_Z \bar{B}_Z$ and the observed distribution of events is beyond the scope of this paper and we leave it to future studies.}. In general, we find that the $H_T$, the $E^{miss}_T$ and the $m_{ll}$ distributions for $G^{*}$ events are similar to those for the GGM model with $\tan\beta=1.5$, gluino mass of 900 GeV and $\mu=$ 600 GeV, shown by ATLAS \cite{Aad:2015wqa}. Most notably, the jet multiplicity distribution of $G^{*} \to B_Z \bar{B}_Z$ events looks closer to that of the observed events compared to the jet multiplicity distributions for the two examples of GGM models considered by ATLAS, which tend to predict a number of signal jets larger than the one observed.

\section{Conclusions and discussions}\label{sec:discussion}

\begin{figure}
\centering
\includegraphics[width=0.48\textwidth]{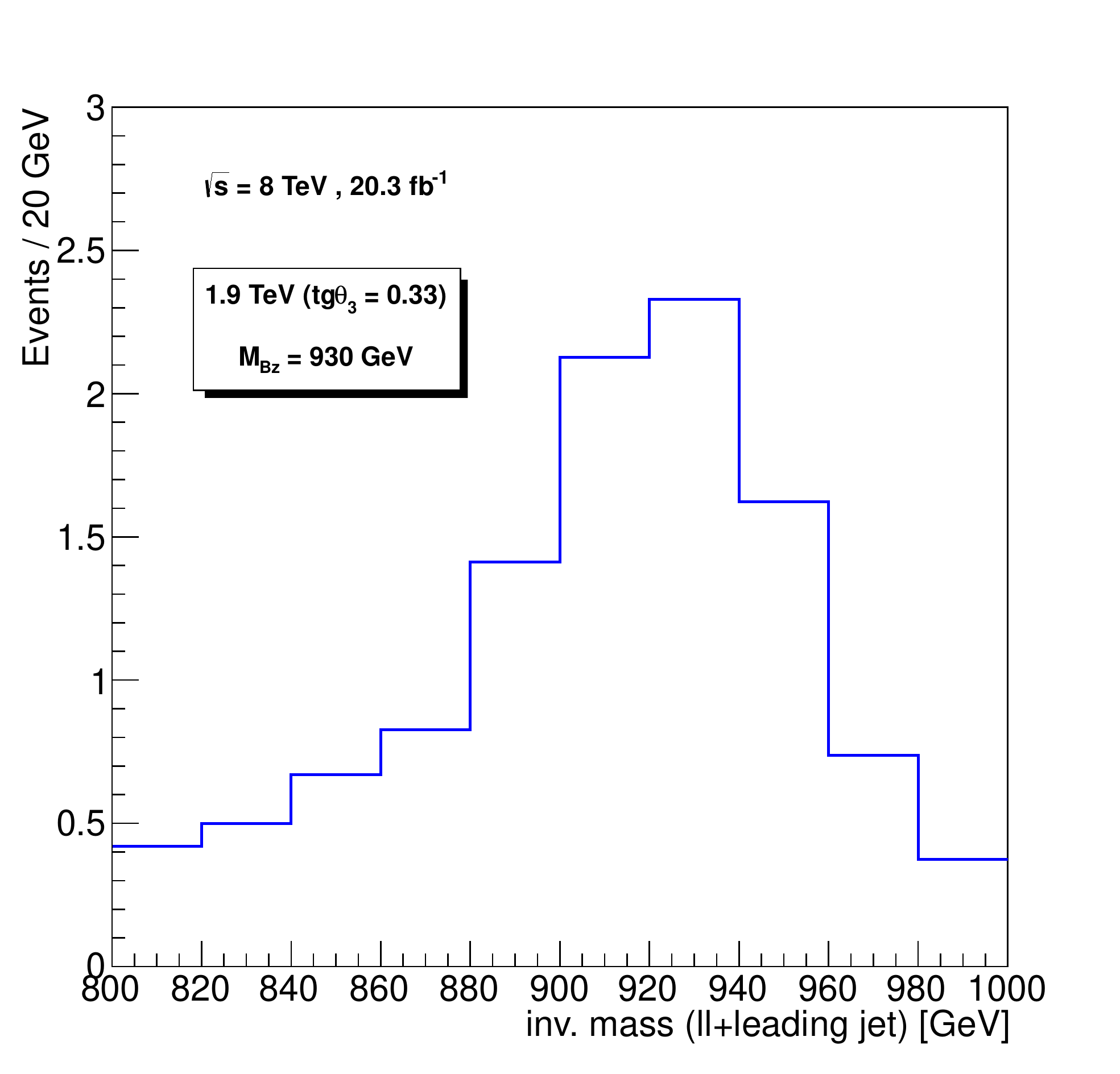}
\includegraphics[width=0.48\textwidth]{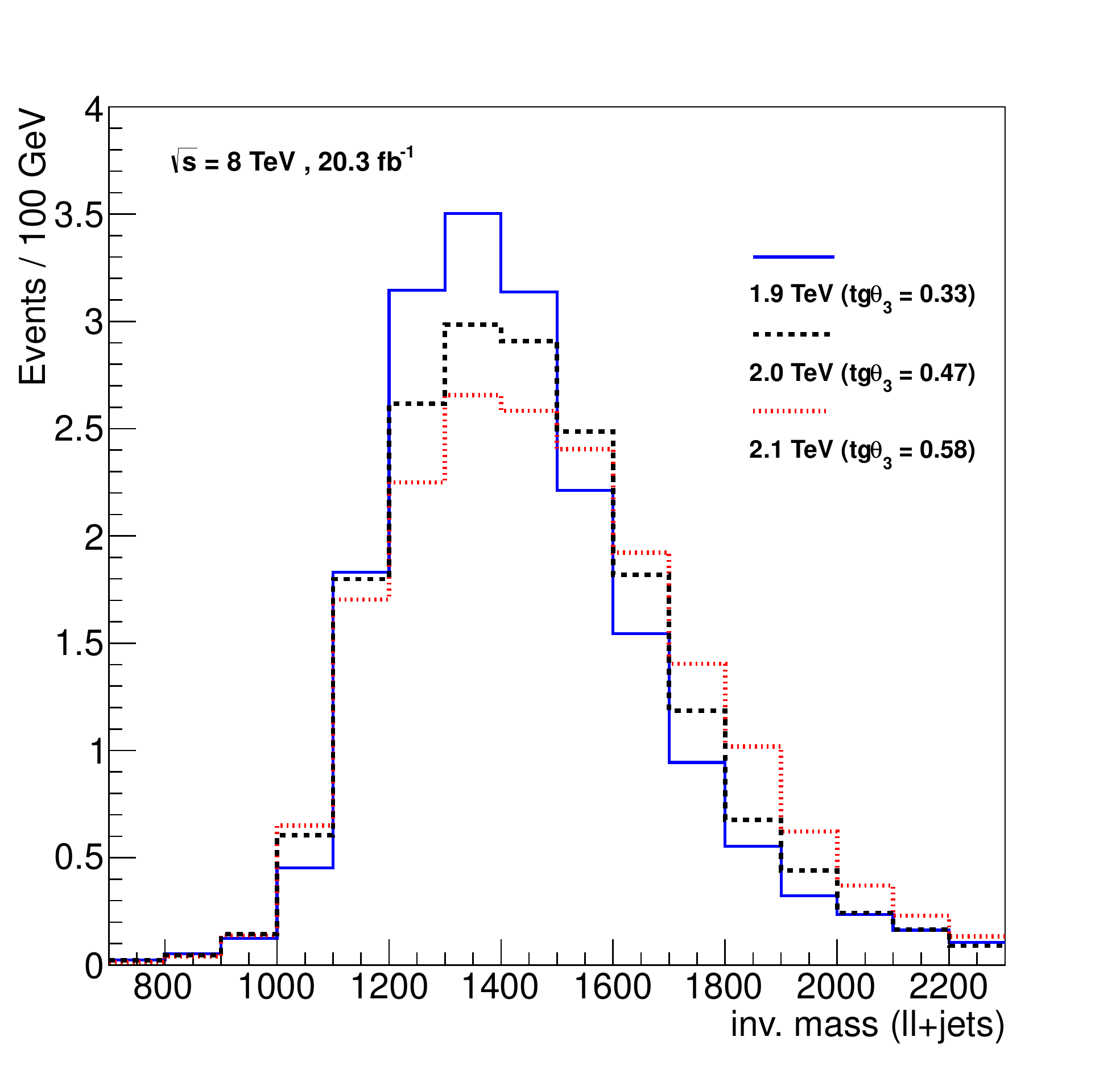}
\caption{\small Event distributions (with 20.3 fb$^{-1}$ at the 8 TeV LHC) for the $G^{*} \to B_Z \bar{B}_Z$ signal with parameters reproducing the ATLAS excess. The vlq mass is fixed at $M_{B_Z}=930$ GeV. Left plot: invariant mass of the two leptons plus the leading jet, for $M_{G^{*}}=1.9$ TeV ($\tan\theta_3=0.33$). Right plot: invariant mass of the two leptons plus all of the signal jets, for $M_{G^{*}}=1.9, 2.0, 2.1$ TeV ($\tan\theta_3=0.33, 0.47, 0.58$).}
\label{fig:MBz-Mkkg}
\end{figure}

In this study we have found that the 3 sigma excess recently measured by ATLAS in events with same flavor opposite sign lepton pairs, peaked on the Z mass, plus jets and large missing energy can be interpreted in composite Higgs models as a result of heavy gluon decays to vector-like quarks. We have analyzed a concrete model where, as an effect of a custodial symmetry protection to the $Zbb$ coupling, bottom partner vector-like quarks, produced in pairs from the heavy gluon decays, contribute significantly to $hhbb$ and $ZZbb$ final states. This latter $ZZbb$ signature, in particular, if one of the $Z$ decays leptonically and the other to neutrinos, can explain the ATLAS excess. Our results are summarized by fig. \ref{fig:results} where we show that the ATLAS excess can be reproduced by the composite Higgs model, in an experimentally allowed region of the parameter space, for heavy gluon masses roughly in a range 1.87 - 2.15 TeV and for heavy gluon couplings to light quarks within $\sim(0.3-0.65) g_S$.\\
An immediate reaction 
to this finding is looking for confirmations of the composite Higgs interpretation of the ATLAS results, especially to distinguish the composite Higgs hypothesis from alternative supersymmetrical explanations. The s-channel exchange of the heavy gluon and 
the two vector-like quark resonances, one of which can be fully reconstructed, are the distinctive features of the composite Higgs signal. Indeed, a simple test of the composite Higgs interpretation could be easily realized by analyzing, for example, the event distributions of the following observables: the invariant mass of the two leptons 
plus the leading jet, which shows a peak around the vector-like quark mass, as shown on the left plot of fig. \ref{fig:MBz-Mkkg}, or the invariant mass of all of the observed objects, leptons and signal jets, of the final state, which presents a kinematic edge\footnote{because of the missing energy, it is not possible to fully reconstruct the heavy gluon resonance.} at high mass, as a result of the heavy gluon exchange; the corresponding plot is shown in fig. \ref{fig:MBz-Mkkg}.   \\
If the ATLAS excess is really due to a composite Higgs/RS theory, further evidences should manifest soon at the upcoming LHC run. The bottom and top partners of the model should be indeed observed in searches for pair production of vector-like quarks. An evidence for the electroweak singlet $\tilde{T}$ could also appear in searches for top-partner electroweak single production; dedicated analyses have been performed in \cite{Vignaroli:2012nf, Andeen:2013zca, Agashe:2013hma}\footnote{with the choice of parameters in eq. (\ref{eq:param}), the electroweak coupling of the $\tilde{T}$ is $\lambda_{\tilde{T}}\simeq 1.4$. For such a coupling the $\tilde{T}$ should be discovered with 100 fb$^{-1}$ at the 14 TeV LHC for masses up to $\sim 1$ TeV \cite{Vignaroli:2012nf}.}.
In general, the results of this study reinforce the importance of focusing on searches for vector resonances which include the decays to vector-like quarks \cite{Bini:2011zb, Vignaroli:2014bpa, Chala:2014mma, Greco:2014aza}. A large portion of the parameter space for composite Higgs models is indeed not accessible to standard search channels, as those in the dijet or $t\bar{t}$ final state.

\section*{Acknowledgments}
This material is based upon work supported by the National Science Foundation under Grant No. PHY-0854889.

\appendix
\section*{Appendix}
In this appendix we demonstrate that $BR(B_Z \to Zb)=BR(B_H \to hb)=1$. This result can be easily obtained from the Yukawa Lagrangian, which in the elementary-composite basis reads \cite{Vignaroli:2012si}:
\begin{equation}
\mathcal{L}^{YUK}= Y_{*} \text{Tr}[\bar{\mathcal{Q'}}\mathcal{H}]\tilde{B} +\cdots \ ,
\end{equation}
where we have written only the terms relevant for our purpose.
$\mathcal{H}$ is the Higgs matrix that, written in terms of the Higgs and Goldstone bosons, is
\begin{equation}
\mathcal{H}=\left( \begin{array}{cc} h-i\it{z} & \sqrt{2}\it{w}^{+}  \\ -  \sqrt{2}\it{w}^{-} &  h+i\it{z} \end{array} \right)=(\bf{1,2,2})_{2/3} \ .
\end{equation}
After the diagonalization of the elementary-composite mixing, which leads in particular to the new $\tilde{B}$ eigenstate:
\begin{equation}
\Big\{\begin{array}{l} \tilde{B}_R= c_{bR} \tilde{B}^{comp}_R + s_{bR} b^{ele}_R \\  b_R= -s_{bR} \tilde{B}^{comp}_R + c_{bR} b^{ele}_R  \end{array} \qquad \frac{s_{bR}}{c_{bR}} = \frac{\Delta_{R2}}{M_{\tilde{B}}} \ ,
\end{equation}
\footnote{here $M_{\tilde{B}}$ denotes the bare $\tilde{B}$ mass, before the elementary-composite and the electroweak mixings.} the Yukawa Lagrangian becomes
\begin{equation}\label{eq:Lyuk}
\mathcal{L}^{YUK}= -Y_{*} s_{bR} \left[ \bar{B}_{-1/3_L} \left ( h -i \it{z}\right ) b_R + \bar{B'}_L \left ( h+i \it{z}\right ) b_R \right]+\cdots \ .
\end{equation}
After the field rotation in (\ref{eq:Bz-BH}), $B_{H(Z)}=\frac{1}{\sqrt{2}}[ B_{-1/3}+(-) B']$, we finally have
\begin{equation}
\mathcal{L}^{YUK}= -\sqrt{2} Y_{*} s_{bR} \left[ \bar{B}_{H_L} h b_R -i \bar{B}_{Z_L}  \it{z} b_R \right]+\cdots \ .
\end{equation}
According to the equivalence theorem, which can be safely applied in our case, since we work in a regime $M_{B_H}\sim M_{B_Z} \gg v$, the above equation implies
\begin{equation}
BR(B_Z \to Z_L b)=BR(B_H \to hb)=1 \ .
\end{equation}
We can also notice that, in a non-custodial scenario, with only the $B'$ bottom partner, eq. (\ref{eq:Lyuk}) simply  reduces to:
\begin{equation}
\mathcal{L}^{YUK}= -Y_{*} s_{bR} \left[ \bar{B'}_L h  b_R +i \bar{B'}_L  \it{z} b_R \right]+\cdots \ 
\end{equation}
and thus the $B'$ decay pattern is: $BR(B' \to hb)=BR(B' \to Z_L b)=0.5$.

\end{document}